\documentclass[useAMS,usenatbib]{mn2e}
\usepackage{graphicx}
\usepackage{amsmath}
\usepackage{makecell}

\usepackage{macros}
\usepackage[usenames,dvipsnames,svgnames,table]{xcolor}
\usepackage{hyperref}
\definecolor{darkblue}{rgb}{0.0,0.0,0.3}
\hypersetup{colorlinks,
            linkcolor=darkblue,urlcolor=darkblue,
            anchorcolor=darkblue,citecolor=darkblue}
\usepackage{txfonts}
\DeclareSymbolFont{cmletters}{OML}{cmm}{m}{it}
\DeclareMathSymbol{v}{\mathalpha}{cmletters}{"76}
\newcommand{\RedeclareMathOperator}[2]{\renewcommand{#1}{}\let#1\relax\DeclareMathOperator{#1}{#2}}

\newcommand{\myvec}{\boldsymbol}

\newcommand\simless\lesssim
\newcommand\simgreat\gtrsim

\newcommand{\avg}[1]{\ensuremath{\langle#1\rangle}}
\newcommand{\abs}[1]{\ensuremath{\lvert#1\rvert}}

\newcommand{\zh}{z_{{\rm h}}}

\newcommand{\fracb}[2]{\left(\frac{#1}{#2}\right)}

\title[Simulations of collapsar jets: 3D instabilities]{Relativistic
  MHD simulations of core-collapse GRB jets: 3D instabilities and
  magnetic dissipation}
\author[Omer Bromberg and Alexander Tchekhovskoy]{Omer Bromberg$^{1}$\thanks{E-mail:
omerb@astro.princeton.edu}\thanks{Lyman Spitzer, Jr.\ Fellow}, Alexander Tchekhovskoy$^{3,
4}$\thanks{Einstein Fellow}\\
$^{1}$Department of Astrophysical Sciences, Peyton Hall, Princeton University, Princeton, NJ 08544, USA\\
$^{2}$Departments of Astronomy and Physics, Theoretical Astrophysics
Center, University of California Berkeley, Berkeley, CA 94720-3411\\
$^3$Lawrence Berkeley National Laboratory, 1 Cyclotron Rd, Berkeley, CA 94720, USA}

\begin{document}

\date{Accepted. Received; in original form}
\pagerange{\pageref{firstpage}--\pageref{lastpage}} \pubyear{2015}

\maketitle

\label{firstpage}

\begin{abstract}
  Relativistic jets are a natural outcome of some of the most violent
  and spectacular astrophysical phenomena, such as the core collapse
  of massive stars in gamma-ray bursts (GRBs) and the accretion onto
  supermassive black holes in active galactic nuclei (AGN). It is
  generally accepted that these jets are powered electromagnetically,
  by the magnetised rotation of a central compact object (a black hole
  or neutron star). However, how the jets produce the observed
  emission and survive the propagation for many orders of magnitude in
  distance without being disrupted by current-driven non-axisymmetric
  instabilities is the subject of active debate. We carry out
  time-dependent 3D relativistic magnetohydrodynamic simulations
  of relativistic, Poynting flux dominated jets {that propagate in
  a spherically-symmetric power-law density distribution}. The jets are launched
  self-consistently by the rotation of a strongly magnetised central
  compact object. This determines the natural degree of azimuthal
  magnetic field winding, a crucial factor that controls jet
  stability. We find that the jets are susceptible to two types of
  instability: (i)~a~global, \emph{external kink} mode that grows on
  long time scales and causes the jets to bodily bend
  sideways. Whereas this mode does not cause jet disruption over the
  simulated distances, it substantially reduces jet propagation
  speed. We show, via an analytic model, that the growth of the
  external kink mode depends on the slope of the ambient medium
  density profile. In flat density distributions characteristic
  of galactic cores, an AGN jet may stall, whereas in stellar
  envelopes the external kink weakens as the jet propagates
  outward; (ii)~a~local, \emph{internal kink} mode that grows over
  short time scales and causes small-angle magnetic reconnection and
  conversion of about half of jet electromagnetic energy flux into
  heat. Based on the robustness and energetics of the internal
  kink mode, we suggest that this instability is the main dissipation
  mechanism responsible for powering GRB prompt emission.
\end{abstract}


\section{Introduction}
\label{sec:introduction}

Relativistic jets are ubiquitous among astrophysical systems that
involve accretion onto compact objects, such as gamma-ray bursts (GRBs), active galactic nuclei (AGNs) and
microquasars \citep[see, e.g. a review by][]{2006IJMPA..21.6015L}. It is commonly agreed that
these jets are powered electromagnetically, most likely by the winding
of magnetic field lines that thread a rotating
central compact object \citep{bz77,kom01,tch10a}. The winding generates a Poynting flux dominated outflow, which eventually becomes the jet, at the
expense of the rotational energy of the central engine. 
Although significant progress has been made recently in understanding how jets are formed magnetically
(see, e.g., \citealt{2015ASSL..414...45T} for a review), 
their physics, most notably the stability properties and the energy
dissipation mechanisms, is the subject of active debate.
In the context of non-relativistic jets, the expectation is that
magnetised jets are strongly unstable to current-driven instabilities. For instance,
\citet{2008A&A...492..621M,2009A&A...507.1203M} show that
non-relativistic jets readily develop current-driven, non-axisymmetric, kink ($m=1$)
modes whose properties depend on the conditions at the jet
base. In the core collapse of a massive star, heavily mass-loaded
jets, which are produced
by the rotation of a protoneutron star
\citep{burrows_snjets_2007}, were found to be so strongly unstable to
the kink instability that they were unable to penetrate the star
\citep{2014ApJ...785L..29M}. %

If this instability
were to extend also to the relativistic, highly magnetised jets
 this would have serious
implications on the properties of relativistic jets and the engines
that launch them. In the context of long-duration GRBs \citep{woo93,mac99} it would imply that magnetised jets are unable to break out of the star,
a necessary condition to form a GRB. Therefore, it would mean that GRB jets have to be created \emph{unmagnetised},
making it impossible to power them
by the electromagnetic spindown of the central object.
However, the observed properties of relativistic jets suggest otherwise. For example, the high power observed in GRB and AGN jets significantly challenges the known non-magnetic energy extraction mechanisms available in these objects
\citep[e.g.][]{Phinney82,2013ApJ...766...31K,2014MNRAS.445L...1L}. Moreover, a recently discovered correlation between the jet magnetic field
strength and the accretion disc luminosity in AGN
\citep{2014Natur.510..126Z}, and several surprising features that are seen in
GRB lightcurves and can be naturally produced by magnetic jets \citep{tg14}
strongly support the magnetic origin of the jet power. 

On the other, hand if the magnetic jets were mostly stable it would be difficult to explain the high energy emission radiated from them \citep[see, e.g.,][]{mb09,nlt09}. In a stable jet about half of the energy remains locked in the magnetic form, out to very large distances \citep{tch10b}. 
In such a case internal shocks, which are commonly invoked to explain the observed high energy emission, are
weak \citep{kennel_mhd_crab_1984} and cannot accelerate efficiently the radiating electrons \citep[e.g.,][]{2010MNRAS.407.2501M,nkt11}. Various
alternative dissipation mechanisms have been discussed in this context, including striped wind like magnetic field configurations, which are susceptible to reconnection \citep[e.g.,][]{2002A&A...391.1141D,
  2005A&A...430....1G, 2011MNRAS.413.2031M,mu12}, microphysical
energy dissipation mechanisms
\citep[e.g.,][]{beloborodov_collisional_grb_10,
  2011ApJ...733L..40M}, and effects from intermittent engine \citep[e.g.][]{2011MNRAS.411.1323G}. 
  To date, no universally-accepted mechanism that is capable of efficiently
converting jet magnetic energy into radiation exists. 
The presence of a local current-driven instability in a mildly unstable jet could be sufficient for triggering magnetic dissipation
and powering the observed GRB emission \citep{2003astro.ph.12347L,2008A&A...480..305G,2012MNRAS.422.3092G}.
This motivates a focused, high-resolution study of jet stability in
the context of core-collapse GRB jets. 

Current-driven instability is mainly a 3D effect and can be highly
non-linear. Thus 3D simulations are needed to study it.  The standard
approach for simulating magnetic jets is sending loops of magnetic
field into the computational domain at a fixed rate
\citep[e.g.,][]{mignone_3d_jets_2010,2014ApJ...781...48G}. In this
case, the strengths of the poloidal magnetic field component, $B_p$
(i.e., lying in the plane passing through the jet axis), and azimuthal
component, $B_\phi$ (perpendicular to the plane), are arbitrary.
However, jet stability depends sensitively on the ratio between the
two \citep{2000A&A...355..818A,nlt09,2012ApJ...757...16M}. Thus, the results of such jet
``injection'' simulations reflect a particular (arbitrary) choice for
the injection boundary condition, making it difficult to interpret
them. Lately, \citet{2014MNRAS.443.1532B} showed analytically that
Poynting-dominated jets that form at the center of collapsing stars
are at least marginally stable and can punch through the stellar
envelope without being disrupted by magnetic instability. However, in the absence of a full 3D numerical simulation,  they had to assume a ratio between $B_\phi$ and $B_p$,  
and did not address the question of magnetic energy dissipation.\footnote{They employed a linear analysis and assumed that the strengths of the toroidal and poloidal field components are comparable
(in the fluid frame). Such an assumption is usually attributed to jets
that propagate through a pre-evacuated funnel and is commonly used
in the studies of jet stability \citep[e.g.,][]{nlt09,mb09}.}

\begin{figure}
  \begin{center}
    \includegraphics[width=\columnwidth]{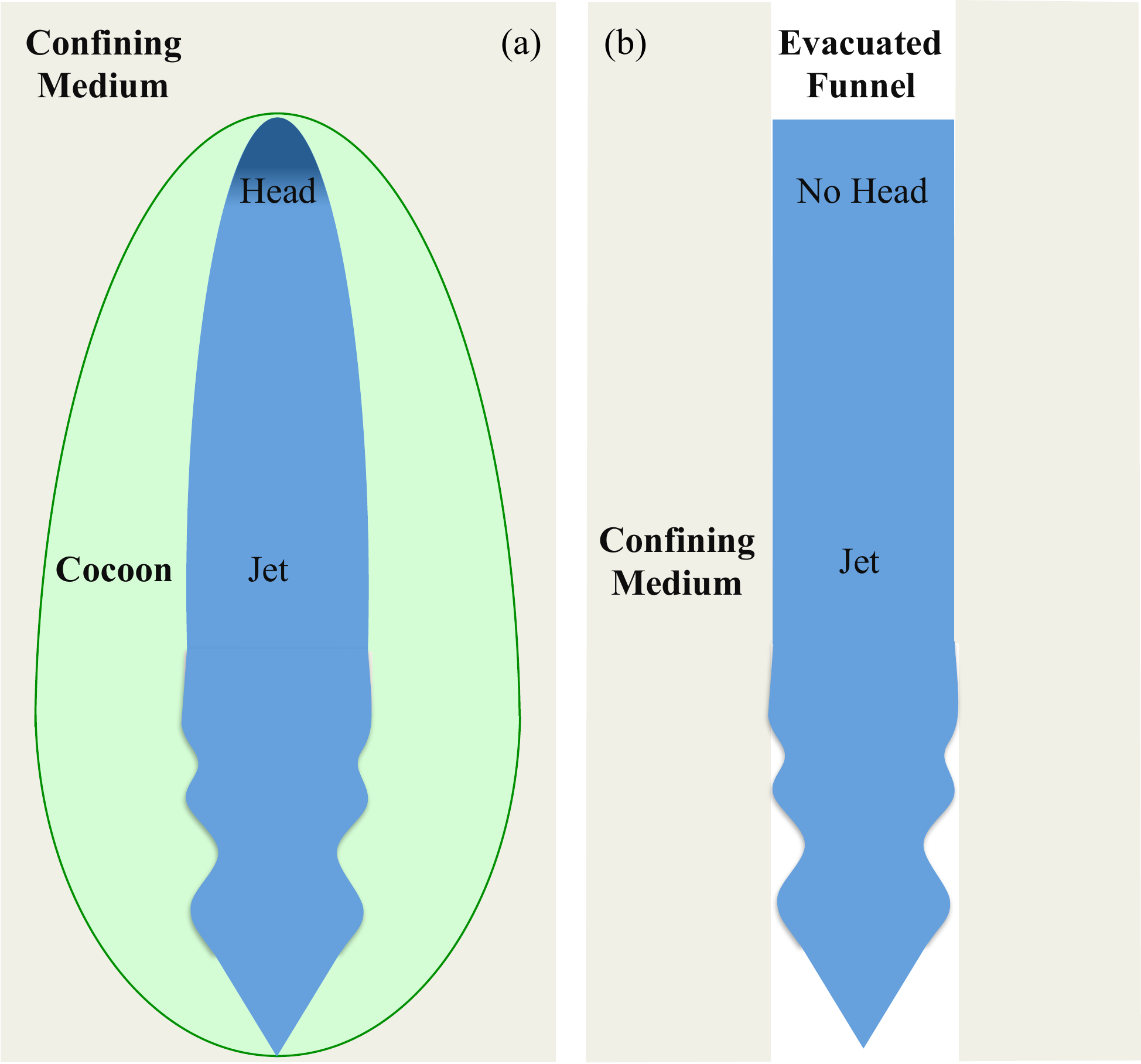}
  \end{center}  
  \caption{A schematic view of a {headed} jet (panel a) that
    propagates in an ambient medium, and a {headless} jet (panel b)
    that propagates in a preexisting evacuated funnel. As the headed
    jet runs into the ambient gas, it slows down, its toroidal
    magnetic flux accumulates and grows in strength, and its outermost
    parts form a head (shown in dark blue), a working surface at which
    the jet drills through the ambient gas and in which the magnetic
    field strength is enhanced. This causes the formation
    of a bow shock (shown with the dark green line), and the shocked ambient gas
    forms the cocoon (shown in green) that collimates the jet into a
    cigar-like shape. In contrast, a headless jet has no ambient gas
    to push through and is free to propagate along a pre-evacuated
    funnel. It propagates faster than headed jets and assumes the
    shape dictated by the funnel and/or ambient pressure
    profile. Because headless jets do not have to push through the
    ambient gas, their toroidal field is weaker and they are more
    stable to kink instabilities than headed jets. }
  \label{fig:headed_vs_headless}
\end{figure}

In nature, the strengths of the poloidal and toroidal magnetic field
components are tightly connected. This is because the jets are produced by the rotation of magnetised compact objects, so $B_\phi$ emerges from the winding of
$B_p$. The winding creates an outward Poynting flux.  In the presence
of an ambient medium,  which confines the electromagnetic outflow, 
the magnetic tension of the azimuthal field builds up,
until it focuses the outflow into twin polar collimated jets
\citep{1996MNRAS.279..389L}, as illustrated in Fig.~\ref{fig:headed_vs_headless}(a).  These jets start propagating once their
pressure becomes high enough to push the ambient medium aside. As the jet propagates, it develops
a slow-moving ``head'': a working surface at which the jet drills
through the ambient material and which is shown in
Fig.~\ref{fig:headed_vs_headless}(a) in dark blue. The head blocks the
free expansion of the toroidal magnetic flux in the jet and keeps the
jet toroidal magnetic pressure high. As a result, the jet pushes
against the head with a greater force.
We term this type of  jets \emph{headed} jets. 
The relative strength of toroidal and poloidal fields in the jet is
therefore linked with the properties of the ambient medium that
collimates it.  Thus, any attempt to analyse jet stability should take
into consideration the presence of the ambient medium and its effect
on the jet magnetic field configuration.  

Jets that expand into a pre-existing, evacuated
funnel are of different nature. They are free to accelerate to super fast-magnetosonic
velocities, as illustrated in
Fig.~\ref{fig:headed_vs_headless}(b). The tip of the latter jets
cannot communicate backward, and the jet material behaves as if it
were part of an infinite jet. In these jets, the toroidal and poloidal
fields can be in equipartition in the fluid frame
\citep{tch08,lyub09,2011PhRvE..83a6302L}. We term these jets
\emph{headless} jets to distinguish them from the {headed} jets
discussed above. 

In this work we address two important questions in the physics of magnetised GRB jets: the stability of the jet as it propagates in the star and the dissipation of its magnetic energy. We present the results from a numerical study of the formation and propagation of relativistic Poynting flux dominated jets in a dense medium typical for a GRB progenitor star. The jets are formed by the rotation of a magnetised neutron star (NS) with a super-strong magnetic field, or a magnetar, in the same manner as described earlier. Thus, the magnetic field configuration is generated
self-consistently, without any ad hoc assumptions about the ratio
between $B_\phi$ and $B_p$. 
This allows us to study jet stability from first principles.
Note that even though we focus on GRB jets, our results are general and  apply to relativistic jets in other astrophysical systems, such as AGN jets.

We start by describing our numerical scheme and the problem setup
(Section~\ref{sec:scheme}) and reviewing the main aspects of kink
instability in magnetically-dominated jets
(Section~\ref{sec:Primer}). We then discuss the difference between
headless jets, which are launched into a preexisting funnel 
(Section~\ref{sec:ideal}), and headed jets, which drill their way though a
dense ambient medium (Section~\ref{sec:2D}). We show that while
headless jets are relatively stable to kink modes, headed jets are
kink-unstable. In Section~\ref{sec:3D}, we analyse the results of our
3D jet simulations and discuss the effect that the kink instability
has on the jets and the limitations of our work. We then construct an analytic model that
describes the jet properties
(Section~\ref{sec:analytical}).  In Section~\ref{sec:astrophysical},
we discuss the astrophysical implications for GRB and AGN jets. We
finish by comparing our results to previous works done on the subject
(Section~\ref{sec:other_works}) and conclude (Section~\ref{sec:disc-concl}). Throughout this paper, we use both
spherical ($r,\theta,\phi$) and cylindrical ($R,\phi,z$) coordinates. 

\section{Numerical Scheme, Units, and Problem Setup}\label{sec:scheme}

We carry out our simulations using the \verb=HARM= code (Gammie et al
2003), with recent improvements (McKinney 2006; Tchekhovskoy et
al. 2007, 2009; McKinney \& Blandford 2009, Tchekhovskoy et al. 2011).
This is a static mesh, 3D, general relativistic magnetohydrodynamic (GRMHD) code
capable of following the evolution of high magnetisation flows while
conserving mass, energy and momentum to machine precision. An
important advantage of the code is its ability to use curved grids, as
we explain below. 

We run both 2D (axisymmetric) and 3D simulations on a grid that spans
a range $(r_{\rm in},r_{\rm out})\times (0,\pi)\times(0,2\pi)$ in
spherical polar coordinates, $(r,\theta,\phi$). We set the boundary
conditions to be reflecting at the poles ($\theta=0, \pi$), free
streaming at the outer radial boundary ($r=r_{\rm out}$), and
periodic in the $\phi-$direction.  The radial grid is uniformly spaced
in $\log r$ out to the radius $r_{\rm br}$, beyond which the grid
becomes progressively sparse.  Tables
\ref{tab:models_grid} and \ref{tab:models_properties} summarise the
parameters of different models we use in this work,
and we give more details in the corresponding sections.

At the inner
radial boundary ($r=r_{\rm in}$), we place a perfectly conducting
sphere carrying a monopole magnetic field. 
At the beginning of the simulation, the sphere is instantaneously spun
up to a constant angular frequency.  This generates Poynting flux
dominated outflow that emanates from the surface of the sphere.  We
spin the inner sphere at a rather large angular frequency
$\Omega \sim0.5 c/r_{\rm in}$ so that the light cylinder, the
cylindrical radius at which co-rotation velocity equals the speed of
light, 
\begin{equation}
R_{\rm L}\equiv \frac{c}{\Omega}, \label{eq:RL}
\end{equation}
is a few times $r_{\rm in}$. This
allows us to speed up the simulation while keeping the physics the
same \citep{kom07}. Outside the sphere we set a static background density that resembles
the interior of a star. The ambient density is a continuous single
power-law distribution $\rho_a=r^{-\alpha}$ that extends between an
inner ``bubble'' of radius $r{_{\rm b}}\geq r{_{\rm in}}$ and the
outer radius of the grid. 
We will focus in this work on $\alpha=2.5$, a value of power-law index
that is characteristic of Wolf-Rayet
stars \citep[e.g.][]{2000ApJ...528..368H}. We assume cold medium,
ignore gravity,\footnote{Whereas gravity and ambient pressure may be
  important close to the jet launching point, they become less
  important at larger distances that are the main focus of this
  work. For instance, the free-fall time scale of the outer parts of
  the star is of order of 100 s, which is much longer than the time it
  takes for the jet to pierce the star, as we will show.} and do not
impart any perturbations to the initial conditions.

Numerical MHD codes cannot evolve vacuum of density. Therefore, we set
the boundary condition at $r=r_{\rm in}$ and the initial density
$\rho$ in the region between $r{_{\rm in}}$ and $r{_{\rm b}}$ by
demanding that the terminal Lorentz factor of the flow, which is
controlled by the initial magnetisation of the flow, is
$\Gamma_{\infty} = \sigma_0 = b^2/4\pi\rho c^2 = 25{-}50$. This value is
high enough ($\gg1$) to lead to a relativistically-magnetised outflow
and at the same time low enough to keep the numerical noise to a
minimum.

In the context of GRBs, we normalise all length scales by $R_{\rm L}=10^7$ cm, the light cylinder radius of a 2 ms NS; the magnetic field strength by
$B_{L}=10^{13}$ G, the magnetic field on the light cylinder; and
all energy and density terms by $B_{L}^2/4\pi$.  Time is measured in the
units of light crossing time of the light cylinder, $R_{\rm L}/c=0.33$
ms. In these units, the dimensionless 3-velocity
$\beta\equiv v/c = z/t$.  
The fiducial value of density,
$\rho_{L,\rm fid}^a=4500$, corresponds to jets of power
$7\times10^{49}\ {\rm erg}\;{\rm s}^{-1}$ %
propagating in a $10M_\odot$ star.
In the context of AGNs, one can use a light
cylinder of $1.5\times10^{13}$ cm and a magnetic field of
$B_L\simeq10^{5}$~G that correspond to 
a maximally rotating supermassive black hole (BH) of mass
$10^8M_\odot$. The fiducial density in these units corresponds to 
$\sim10^{-7}$~g~cm$^{-3}$ at the light cylinder (see Table
\ref{tab:models_properties}). Unless mentioned otherwise, we will
give physical units in the text in the context of GRB jets.

Table~\ref{tab:models_grid} summarises the parameters of the numerical grid
we used to carry out the simulations, while
Table~\ref{tab:models_properties} lists the physical parameters of the
setup, giving them in dimensionless units as well as
physical units specialised to GRBs and AGNs. 
The essential parameters that
control the jet stability are given in the first two columns of Table~\ref{tab:models_properties}: (i) the dimensionless
magnetic field energy density (in units of ambient density),
$\propto B_{\rm L}^2/\rho_{\rm L} c^2$, which is a proxy for jet
power $L_{\rm j}$, and (ii) the power-law index of the ambient density distribution,
$\alpha$. The third parameter,
$\propto L_{\rm j}/B_{\rm L}^2 R_{\rm L}^2$, reflects the fraction of
the stellar surface area that launches the outflow, which is slightly different in
2D and 3D simulations, as we discuss below. The fourth parameter, $\sigma_0$, sets the
initial magnetisation of the outflow and 
does not influence the simulation outcome so long as $\sigma_0\gg1$: 
it affects the results only at larger distances
than those simulated here (e.g., outside the star).

\subsection{2D simulation setup}

\begin{figure}
  \begin{center}
    \includegraphics[width=\columnwidth]{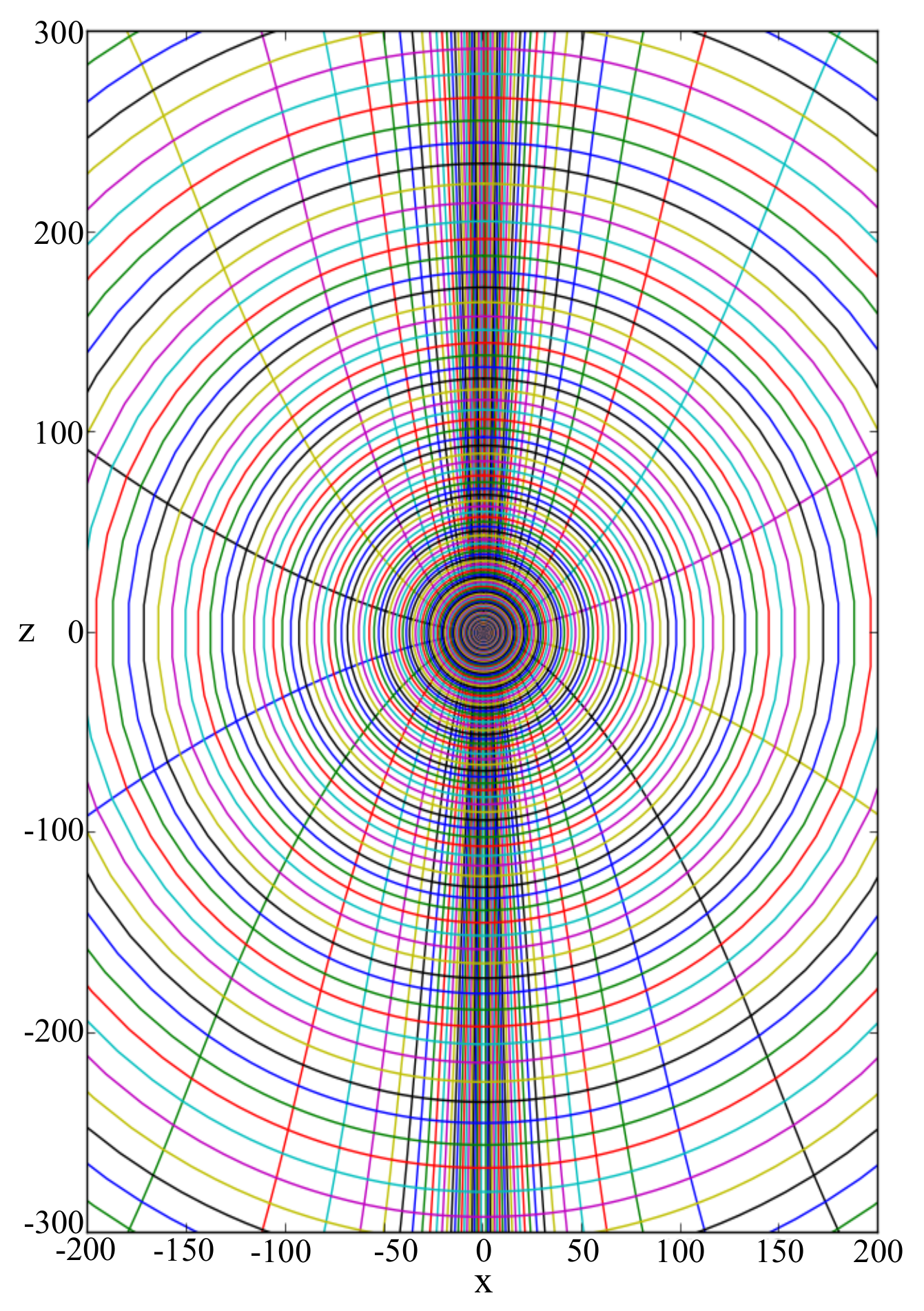}
  \end{center}
  \caption{The shape of the grid lines used in our 2D (axisymmetric) simulations
  (for clarity, we show every fourth grid line).  The grid lines
  collimate toward the rotational axis (the $z-$axis) thereby concentrating the resolution in the body of the jet. In the 3D
  simulations, the collimation is performed toward the $x-$axis to
  avoid the interaction of the jets with the coordinate singularity
  (see Sec.~\ref{sec:3d-simulation-setup}), with the $\theta$ and
  $\phi$ angles of the grid lines deformed independently, in the same way
  $\theta-$lines are deformed in the Figure.}
  \label{fig:grid}
\end{figure}

We carry out our 2D simulations on a spherical polar grid that is
modified to concentrate grid cells toward the polar axis, where the
jets form, by deforming
the radial grid lines into parabolae, as seen in
Fig.~\ref{fig:grid}. We direct the rotational
axis along the polar axis ($\theta=0$) and use a resolution of $1024\times256\times1$ in the $r-$, $\theta-$ and $\phi-$
directions, respectively (see Table \ref{tab:models_grid}). At this resolution the light cylinder
($R_{\rm L}=2r_{\rm in}$) is resolved by $5$ cells, at an altitude
$\abs{z}=1000r_{\rm in}$,  and the jet width by $40$
cells.  We ran the simulations with both dipole and
monopole magnetic field geometries and verified that in both cases the
resultant jet properties are qualitatively the same. This is because
in both cases the jet field lines stretch from the central source to
the jet head and return back to the source (see Section
\ref{sec:2D}). The only essential difference is that in the dipole
case there is a region of closed field lines around the equator of the
central object, which modify the conditions near the base of the jet
\citep{2015arXiv150301467T}. We verified that for an aligned rotator, this effect does not substantially affect
the properties of the jet at higher altitudes. We will therefore
restrict ourselves to the monopole field. This choice
corresponds to the assumption of a large-scale magnetic flux threading
the progenitor star \citep{buc09}.

\label{sec:2d-simulation-setup}

\subsection{3D simulation setup}
\label{sec:3d-simulation-setup}
In the 3D runs we use a slightly modified configuration of grid lines
and magnetic field topology. The existence of a polar coordinate
singularity in the grid along the ${z}-$axis leads to numerical
difficulties when matter and magnetic fields pass through the pole. As a
result, magnetic field lines cannot cross the pole freely and get
wrapped around the $z-$axis. This is not a problem in the 2D case,
since all quantities have azimuthal symmetry, and no mass, momentum,
or energy flows through the pole. However in the 3D case, we found that the imperfect flow through the axis leads to an artificial stability of the jets. 
To avoid this non-physical behaviour, we opted for reorienting the
rotational axis of the central objet to point along the
$x$--direction \citep[see, e.g.,][]{2008A&A...492..621M}, and we collimate the radial grid lines along the
$x$--axis accordingly. Namely, the collimation is performed such that the
$\theta-$ and $\phi-$angles of the radial grid lines deformed independently. Note that when
reporting the simulation results, we will use the natural coordinate
system set by the jet orientation, with the $z-$axis pointing along
the rotational axis. 

To avoid sending any fluxes
through the $z-$axis, we spin not the entire star but only its 
northern and southern polar regions that are within
$70^\circ$ of the rotational axis. The
rotation is uniform within $50^\circ$ of the axis and smoothly tapers off
down to zero at $70^\circ$.  We run 3D
simulations at two resolutions: a fiducial resolution of
$128 \times 96 \times 192$ cells in the $r-$, $\theta-$, and
$\phi-$directions, respectively, and a high resolution, for which the
resolution is doubled in all $3$ dimensions (see Table~\ref{tab:models_grid}).

\begin{table}
\begin{center}
  \caption{Simulation grid parameters for the various models we present in
    this paper. See Sec.~\ref{sec:scheme} for the description of the
    parameters in the table.}
 \begin{tabular}{@{}l l c c c c @{}} 
 \hline
 \thead{Model \\ name} & \thead{Resolution\\ ($N_r\times N_\theta\times N_\phi$)} & \thead{$\displaystyle \frac{r_{{\rm in}}}{R_{{\rm L}}}$} & \thead{$\displaystyle\frac{r_{{\rm b}}}{R_{{\rm L}}}$} & \thead{$\displaystyle\frac{r_{\rm br}^{\star}}{R_{{\rm L}}}$} & \thead{$\displaystyle\frac{r_{\rm out}}{R_{{\rm L}}}$}\\ [0.5ex] 
 \hline\hline
 M2 &  $1024\times256\times1$ & 0.5 & 2 &$500$ &$5000$ \\ 
 M2Cyl & $1024\times256\times1$ & 0.5 & 2 &$500$ &$5000$ \\
 M3 & $128\times96\times192$ & 0.8 & 0.8 & $800$&$8000$ \\
 M3HR & $256\times192\times384$ & 0.8 & 0.8 &$800$ & $160$  \\
 M3LP & $128\times96\times192$ & 0.8 & 0.8 &$800$ &$8000$ \\  
 \hline
 \multicolumn{6}{l}{}\\
 \multicolumn{6}{l}{$^\star$ The radius beyond which the radial grid becomes}\\
 \multicolumn{6}{l}{\phantom{$^\star$}  progressively sparse.}\\
\end{tabular}\label{tab:models_grid}
\end{center}
\end{table}

\begin{table*}
\begin{center}
  \caption{The physical parameters of the various models we present in
    this paper. The two essential parameters describing the astrophysical system
    are the first two ones: the dimensionless magnetic field energy
    density (in units of ambient density) and the ambient density
    slope. See Sec.~\ref{sec:scheme} for the description of the
    parameters in the table. Unless otherwise noted, cgs units are
    used.}
 \begin{tabular}{@{}l c c c c@{\qquad}c@{\;}c@{\;}c@{\;}c@{\qquad}c@{\;}c@{\;}c@{\;}c@{\;}c@{}} 
 \hline
 & \multicolumn{4}{c}{Scale-free parameters} & \multicolumn{4}{l}{GRB-scaled par.}& \multicolumn{5}{c}{AGN-scaled parameters}\\
 \thead{Model \\ name} &\thead{$\displaystyle \frac{4\pi\rho_{\rm L}c^2}{B_{\rm L}^2}$}  & $\alpha$& \thead{$\displaystyle \frac{3L_{\rm j}}{B_{\rm L}^2R_{\rm L}^2c}$$^\dagger$}& \thead{$\displaystyle \sigma_0$$^\ddagger$}& \thead{$\displaystyle\frac{P}{\rm ms}$} & \thead{$\displaystyle\frac{R_{\rm L}}{10^7}$}& \thead{$\displaystyle\frac{L_{\rm j}^\diamondsuit}{10^{49}}$}& \thead{$\displaystyle \frac{M}{M_\odot}$}& \thead{$a_{\rm BH}$}& \thead{$\displaystyle \frac{M_{\rm BH}}{10^8M_\odot}$} & \thead{$\displaystyle\frac{R_{\rm L}}{10^{13}}$} & \thead{$\displaystyle\frac{L_{\rm j}^\diamondsuit}{10^{46}}$}& \thead{$\displaystyle\frac{\rho_{\rm L}}{10^{-7}}$}\\ [0.5ex] 
 \hline\hline
 M2 &  4500 & 2.5 & 1 & $50$ & $2$ &  $1$ & $7$ & 10 & $1$ & $2.5$ & $1.5$ & $2.3$ & $3$\\ 
 M2Cyl &         $10^5$ & 0   & 1 & $50$ & $2$ & $1$ &  $7$ & 10 & $1$ & $2.5$ & $1.5$ & $2.3$& $3$\\
 M3 &          4500 & 2.5 & 0.76 & $25$ & $2$ & $1$ & $5$ & 10 & $1$ & $2.5$ & $1.5$ & $1.7$& $3$\\
 M3HR    &    4500 & 2.5 & 0.76 & $25$ & $2$ & $1$ & $5$ & 10 & $1$ & $2.5$ & $1.5$ & $1.7$& $3$\\
 M3LP & 45000 & $ 2.5 $ & 0.76 & $25$ & $2$ & $1$ &  $0.5$ & 10 & $1$ & $2.5$ & $1.5$ & $0.17$& $3$\\  
 \hline
  \multicolumn{11}{l}{}\\
  \multicolumn{11}{l}{$^\dagger$ The total Poynting flux from a rotating monopole field is $2B_L^2R_{\rm L}^2c/3$}\\
  \multicolumn{11}{l}{$^\ddagger$ Defined as $b^2/4\pi w$ at the base
   of the jet, where $w\equiv\rho+u+p$ is the enthalpy density.}\\
 \multicolumn{11}{l}{$^\diamondsuit$ $L_{\rm j}$ denotes the power of a single jet.}\\
\end{tabular}\label{tab:models_properties}
\end{center}
\end{table*}

\section{Jet stability and acceleration}\label{sec:Primer}

Magnetically-dominated flows are subject to various types of
instabilities. In a narrow jet, the fastest growing instability is the
kink ($m=1$) instability~\citep{beg98,lyu99}. It excites large scale helical motions in
the jet and can lead to the dissipation of the magnetic
energy or even the disruption of the entire jet. 
Here and below, we will use
$\myvec{b}$ for the magnetic field in the proper fluid fame
and $\myvec{B}$ for the magnetic field in the lab frame.  

The instability evolves on a time scale that is of the order of the
Alfv\'{e}n travel time around the unstable region.  Suppose a jet consists of an inner core, which is located at $R\le R_0$ and dominated
by the poloidal field, $\myvec b_p\equiv b_R \myvec{\hat e_R} + b_z
\myvec{\hat e_z}$, and an outer region, which is dominated by the toroidal field, $b_\phi$. Here, $b_p$ and $b_\phi$ are the poloidal and toroidal field
components measured in the proper frame, respectively. 
By construction, the pitch
angle of the magnetic field in the proper frame, $b_p/b_\phi$, is of order
unity at the edge of the core. Then, the growth
timescale of the fastest growing kink mode in the lab frame is
\citep[e.g.][]{2000A&A...355..818A}:
\begin{equation}\label{eq:t_kink}
   t_{\rm kink}\simeq \frac{2\pi R_0\gamma_{\rm j}}{v_A}\frac{b_p}{b_\phi},
\end{equation}
where $\gamma_{\rm j}$ is the bulk Lorentz factor of the
flow and $v_A$ is the Alfv\'{e}n velocity, 
\begin{equation}\label{eq:v_A}
\frac{v_{\rm A}}{c} = \frac{b}{\sqrt{b^2+4\pi w}},  
\end{equation}
with $b=({b_\phi^2+b_p^2})^{1/2}$ being the total field strength in the fluid frame, $w\equiv \rho+u+p$ the plasma enthalpy, and $u$ and $p$ gas internal energy and pressure, respectively. 

From eq.~\eqref{eq:t_kink} we see that strong toroidal
fields tend to destabilise the jet and lead to the growth of kink
instability, whereas strong poloidal fields tend to stabilise the jet
against kinking.
Typically, it takes about $5{-}10$ growth times for the instability to
evolve to  sufficiently affect the global structure of the magnetic
field.  \citet{2012ApJ...757...16M} found that the instability growth
time is linked with the transverse profile of the fluid frame
toroidal magnetic field. Kink modes grow faster in steep profiles e.g. $b_\phi\propto R^{-1}$, while in shallower profiles the instability
takes longer to grow. Relativistic motions increase the
growth time in the lab frame even further due to the time dilation, as evident in eq.~\eqref{eq:t_kink}, and therefore increase the stability against kink modes even further.

The fact that the growth of the kink modes requires Alfv\'{e}n waves to travel several times around the jet, implies that fluid elements must be able
to communicate efficiently across the jet; otherwise the instability
cannot grow. In a highly magnetised flow, with $b^2\gg4\pi w$, we have
$v_A\simeq c$ (eq. \ref{eq:v_A}), for which the condition for strong causal contact is \citep{1992SvAL...18..356L}:
\begin{equation}
  \gamma_{\rm j}\theta_j\lesssim1.
\label{eq:gammatheta}
\end{equation}
Equation~\eqref{eq:gammatheta}
states that a fluid element with a Lorentz factor $\gamma_{\rm j}$ moving
along a field line that makes an angle $\theta_j$ with the jet axis, can
communicate with the jet axis in a time that is shorter than the
time it takes it to double its altitude. Only those regions in the jet
that are strongly causally connected with the axis can become kink unstable
\citep{1992SvAL...18..356L,2014arXiv1408.3318P}. 

\section{Axially symmetric headless jets: jets moving through low-density medium}\label{sec:ideal}

It is convenient to study highly magnetised jets in the force-free approximation \citep[e.g.][]{beszak04,nar07,lyub09,tch08,2014MNRAS.443.1532B}. In this approximation, gas inertia is neglected, and the motion of the plasma is given by the drift velocity of the electromagnetic field \citep{tpm86,bes97,nar07},
\begin{equation}\label{eq:v_d}
  \beta_{\rm d} = \left|\frac{\myvec{E}\times\myvec{B}}{B^2}\right|=\frac{E}{B},
\end{equation} 
where $\myvec E$ is the electric field vector in the lab frame.  In
this section we consider \emph{headless} jets that are free to move along
a pre-existing evacuated funnel. In such jets, the lab-frame toroidal
magnetic field can be approximated as \citep{bes97,nar07,tch08},
\begin{equation}\label{eq:Bphi_infinite}
  -B_\phi\simeq E=\Omega RB_p/c,
\end{equation}
where $\Omega$ is the angular frequency of the central object, and
$B_p\equiv(B_R^2+B_z^2)^{1/2}$ and $B_\phi$ are the lab-frame magnetic field strengths of the poloidal and toroidal components, respectively. Plugging 
eq.~\eqref{eq:Bphi_infinite} into
eq.~\eqref{eq:v_d}, we obtain the Lorentz factor, 
\begin{equation}\label{eq:gamma_d}
  \frac{1}{\gamma_{\rm j}^2}=(1-\beta_{\rm d}^2)=\frac{B_p^2+B_\phi^2-E^2}{B^2}\simeq\frac{b_p^2}{B^2}+\frac{b_\phi^2}{B^2},
\end{equation}
where we approximate $B_p\simeq b_p$ and
$ b_\phi^2\simeq B_\phi^2-E^2$, since the motions are mostly in the poloidal direction.  In the limit $b_\phi\simless b_p$ we can approximate:
\begin{equation}\label{eq:g_j_headless}
  \gamma_{\rm j}\simeq \sqrt{1+\frac{B_\phi^2}{B_p^2}} 
  \simeq \sqrt{1+\frac{R^2}{R_{\rm L}^2}} \approx \frac{R}{R_{\rm L}}.
\end{equation}
The last approximate equality holds in the limit
$R\gg R_{\rm L}$, in which $\gamma_{\rm j}$ scales linearly with the
cylindrical radius, similar to the acceleration in a hydrodynamic jet
\citep{1993MNRAS.263..861P}. The approximation $b_\phi\simless b_p$ which leads to eq.~\eqref{eq:g_j_headless} holds
close to the jet axis, where the plasma maintains strong causal
connection ($\gamma_{\rm j}\theta_j\lesssim1$).  If a jet is continuously
collimated, the external collimating forces need to be communicated
across the jet. This requires the jet to maintain a
strong causal contact with the axis. Therefore, the linear growth of
$\gamma_{\rm j}$ can only be maintained as long as
$\gamma_{\rm j}\theta_j\approx (R/R_{\rm L})\times \theta_j \simless 1$. At larger values of $\theta_j$, the 
requirement that the external collimating force  should be
communicated across the jet cross-section limits the Lorentz factor
to a value given by eq.~\eqref{eq:gammatheta}, i.e.
$\gamma_{\rm j} \simeq 1/\theta_j$ \citep{tch08,tch09,lyub09}. 

We now turn  to the evaluation of jet stability to kink modes. In a
steady state, there is a force balance in the transverse direction, at
the jet inlet, between the magnetic pressure force and
the toroidal and poloidal hoop stresses, expressed via the following force-balance
equation:
\begin{equation}\label{eq:dbdr}
  \frac{db^2}{8\pi dR}+\frac{b_\phi^2}{4\pi R}+\frac{E^2-B^2_p}{4\pi R_{\rm curv}}=0,
\end{equation}
where the first term is the magnetic pressure gradient, the second is
the hoop stress of the toroidal field and the third is the hoop stress
of the poloidal field, with $R_{\rm curv}$ denoting the poloidal curvature
radius of magnetic field lines (see also \citealt{tch08}).  A jet
cannot be self-collimated and needs to be pressure supported on the
outside by some external confining force
\citep{eichler_93,1994PASJ...46..123T,begelman_asymptotic_1994,bes98}.
In a jet that expands into a preexisting evacuated funnel, the force
is dictated by the shape of the funnel walls. On the other hand, if
the jet is pushing its way through an ambient medium, the confining
force comes from a pressurised cocoon that is formed as a result of the jet propagation
(see Fig.~\ref{fig:headed_vs_headless} and Sec.~\ref{sec:2D}). The longitudinal (along the jet) pressure
profile in the cocoon is relatively uniform and collimates
the jet into a cylinder.
This implies that the
poloidal field lines are mostly along the $z-$direction, their curvature
radius is very large, $R_{\rm curv}\gg R$, and the third term in
eq.~\eqref{eq:dbdr} is negligible in comparison with the rest of the terms.

In order to solve eq.~\eqref{eq:dbdr}, we need to know the
relationship between $b_p$ and $b_\phi$. Far from the jet axis, the velocity is mostly poloidal, and we can approximate $b_p\simeq B_p$ and
\begin{equation}\label{eq:b_phi_b_p_inf}
  b_\phi\simeq \frac{B_\phi}{\gamma_{\rm j}} \simeq
  \left\{
\begin{array}{ll}
 \displaystyle\fracb{R}{R_{\rm L}}b_p & \mbox{, $R\ll R_{\rm L}$}, \\
 b_p & \mbox{, $R\gg R_{\rm L}$},
\end{array}
\right.
\end{equation}
where we used the eqs.~\eqref{eq:Bphi_infinite} and \eqref{eq:g_j_headless} to
connect $B_\phi$ and $B_p$. Thus, inside the light cylinder the
magnetic field is dominated by the poloidal field, while outside the
light cylinder the two are in equipartition, $b_p\simeq b_\phi$.  Substituting this into
eq. \eqref{eq:dbdr}, we get that the profile of the toroidal field
follows
\begin{equation}\label{eq:b_phi_inf}
  b_\phi\propto
  \left\{
\begin{array}{ll}
 R & \mbox{, $R\ll R_{\rm L}$}, \\
R^{-1/2}& \mbox{, $R\gg R_{\rm L}$}.
\end{array}
\right.
\end{equation}
Such a flat configuration of toroidal field outside $R_{\rm L}$, was
shown by \citet{2012ApJ...757...16M} to suppress the growth rate of
kink modes. Moreover, the larger Lorentz factors decrease the lab-frame
growth rate even further and result in a largely stable jet (see eq.~\ref{eq:t_kink}). Making
use of eq.~\eqref{eq:b_phi_b_p_inf},  we obtain for the poloidal field,
\begin{equation}\label{eq:b_p_inf}
  b_p\propto
  \left\{
\begin{array}{ll}
 R^0 & \mbox{, $R\ll R_{\rm L}$}, \\
R^{-1/2}& \mbox{, $R\gg R_{\rm L}$}.
\end{array}
\right.
\end{equation}

\begin{figure}
\begin{center}
\includegraphics[width=\columnwidth]{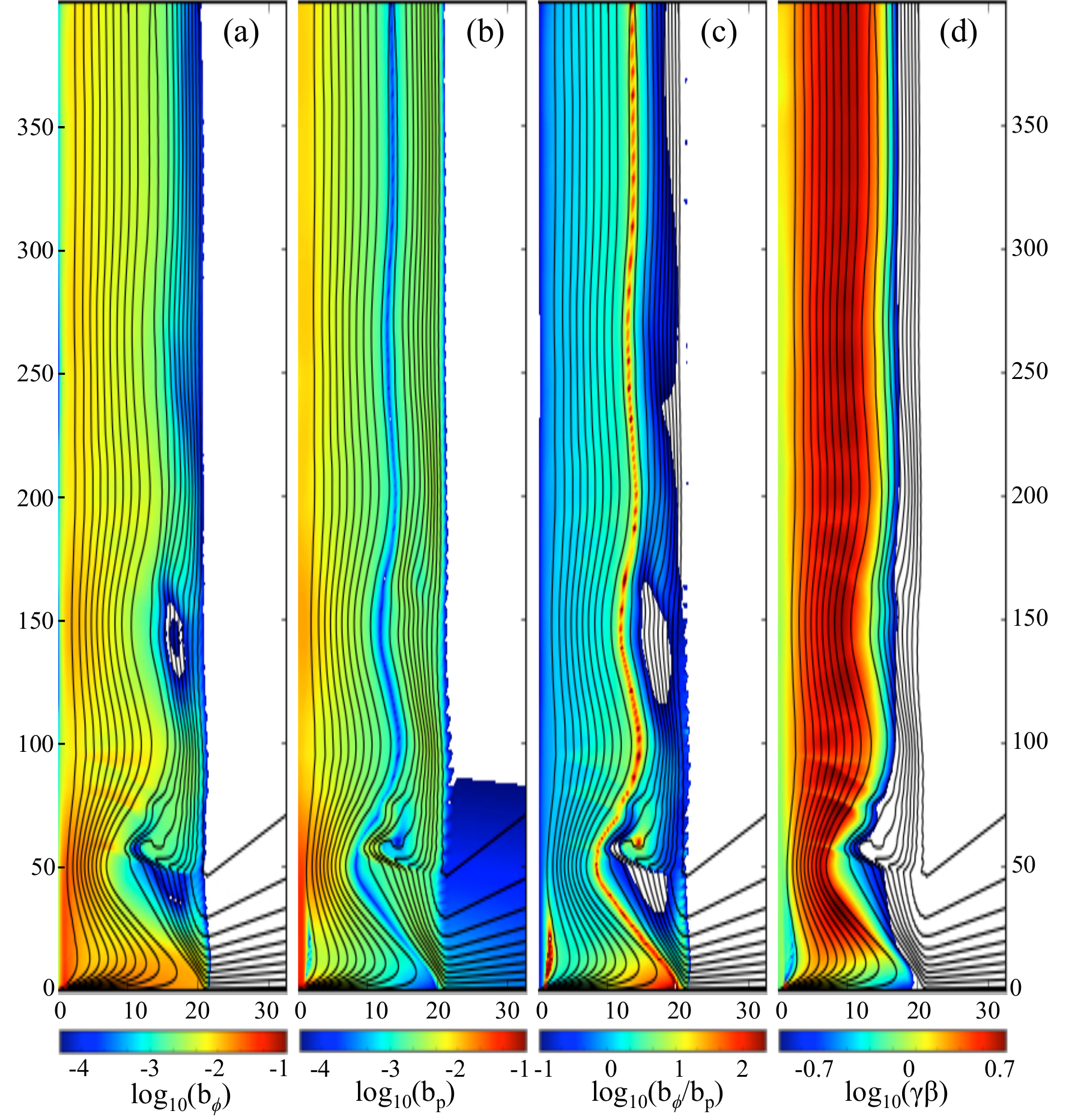}
\end{center}
\caption{The 4 panels show, from left to right, the values of
    $b_\phi$, $b_p$, $b_{\phi}/b_p$ and $\gamma\beta$ in
    a headless 2D jet. The unit of length is the
    light cylinder radius, $R_{\rm L} = 1$. The jet propagates inside of a preexisting
    low-density cylindrical funnel (of radius $R = 20R_{\rm L}$) carved out in a high-density
    ambient medium. The central object produces
  an equatorial outflow, which bounces off the funnel walls and
  re-collimates at $z\sim10$, forming a pair of jets (we show only one
  of them). After several oscillations, the
  cylindrical radius of the jet settles to $R_{\rm j}\sim13$. The
  jet is surrounded by returning magnetic field lines, which fill
  the rest of the cavity but carry very little to no energy flux, as
  indicated in the leftmost panel by the low value of $b_\phi$
  there. The two middle panels show that in the jet the toroidal and poloidal
  magnetic field components are roughly in equipartition in the fluid
  frame, $b_\phi\lesssim
  b_p$. The rightmost panel shows that the jet Lorentz factor is
  largest at the edge of the jet, as expected from the analytic models
  (see Sec.~\ref{sec:ideal}).  See Fig.~\ref{fig:2D_profiles_headless}
  for the transverse profiles of quantities shown in this
  figure. Since the ambient density is extends out to infinity and the walls are rigid, the headless jet eventually reaches a steady state. See
  Figs.~\ref{fig:2D_p_rho_jets}--\ref{fig:2D_profiles_headed} to compare to headed jets
  that do not have a pre-existing funnel to propagate through and have
  to drill one themselves.}
\label{fig:2D_headless_panels}
\end{figure}

\begin{figure}
\begin{center}  
    \includegraphics[width=\columnwidth]{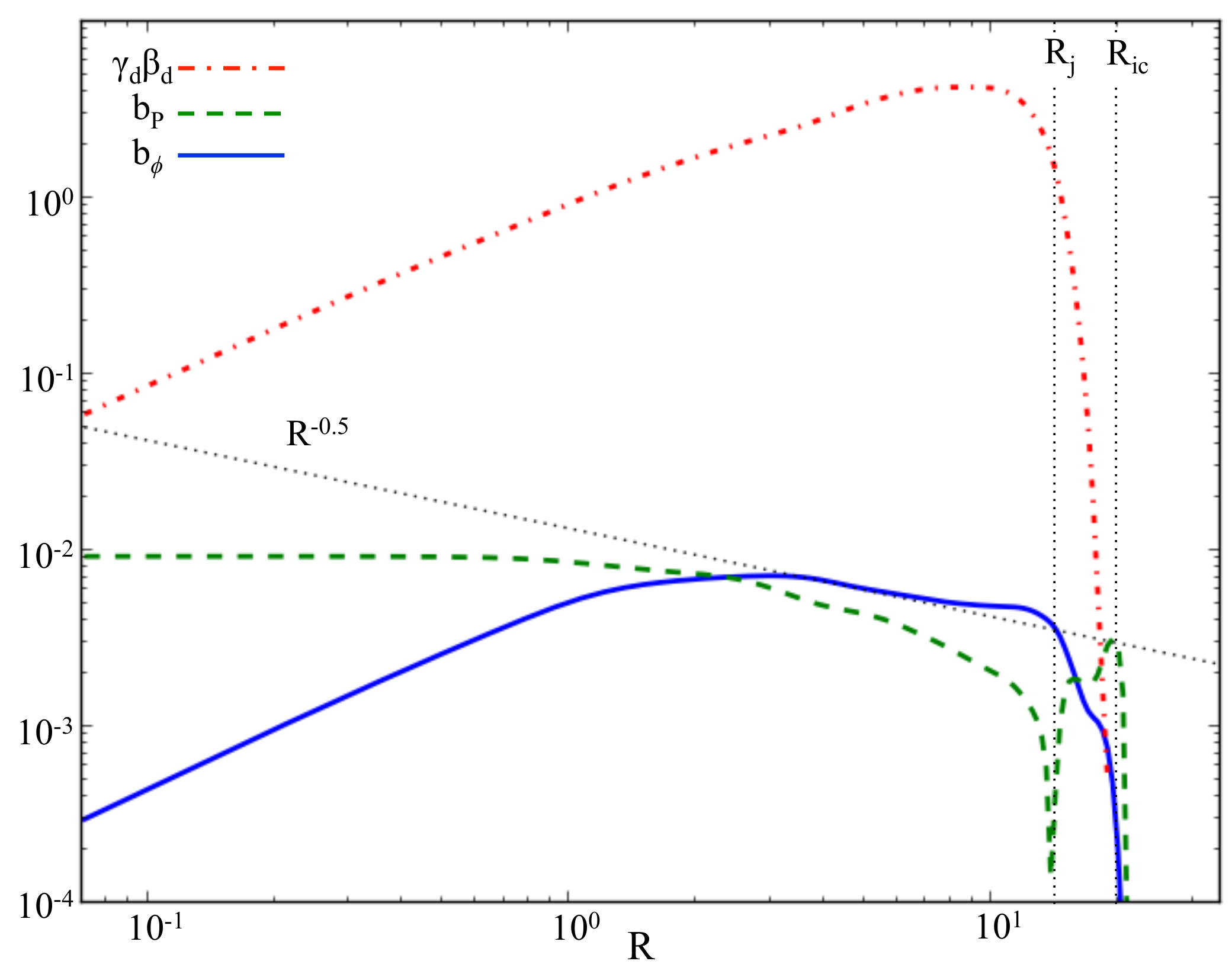}
\end{center}
\caption{The  cross-sectional profiles of the drift 4-velocity (dash-dotted red),
  $b_p$ (dashed green)  and $b_\phi$ (solid blue) in a headless jet (model M2Cyl),
  taken at an altitude $z=200R_{\rm L}\sim2\times10^{9}$~cm.
  The fluid-frame toroidal magnetic field increases inside the light cylinder, $R_{\rm L}=1$,
  peaks just outside of it and then drops off as $b_{\phi}\propto
  R^{-1/2}$. The profile of $b_p$ is flat inside and 
  slowly decreases outside of the light cylinder until the jet edge at
  $R_{\rm j}=13R_{\rm L}$. The sharp decrease in $b_p$ at $R_{\rm j}$ occurs due to the
  flip in the direction of the field lines at the edge of the jet,
  marked by $R_{\rm j}$. The returning field lines fill the rest of the
  cavity up to $R_{\rm ic}=20R_{\rm L}$ and form an inner cocoon that separates
  the jet from the medium outside the funnel. The poloidal and
  toroidal magnetic field components are roughly in equipartition inside the jet, $b_p\simeq b_\phi$. As a  result (see eq.~\ref{eq:g_j_headless}), the drift velocity $\gamma_{\rm d}\beta_{\rm d}$ scales linearly with $R$,
  and reaches a peak value of $\gamma_{\rm d}\beta_{\rm d}=5$ at $R=7R_{\rm L}$. The full morphology of the jet is shown in Fig.~\ref{fig:2D_headless_panels}.}
\label{fig:2D_profiles_headless}
\end{figure}

To illustrate the properties of headless jets, we run a 2D simulation
with a rotating sphere threaded with a monopole magnetic field.  The
sphere is placed in an empty cylindrical funnel with a radius $R=20$,
carved out in a very dense ambient medium (see
Table~\ref{tab:models_properties}), where the medium plays the role of
a rigid, confining wall. The funnel radius is set to be $R\gg R_{\rm L}$ to
allow for relativistic motions of the jet material. As we discussed in
Section ~\ref{sec:introduction}, the rotation of the central sphere
generates toroidal field that expands outward as an Alfv\'{e}n
wave. The expansion is blocked by the confining medium, resulting in a
buildup of toroidal tension which collimates the poloidal field lines
toward the rotational axis (see Fig.~\ref{fig:headed_vs_headless}b).
The field lines undergo an initial oscillatory phase and relax into a
cylindrical shape along the evacuated funnel (see
\citealt{2005MNRAS.357..918B} for a similar effect in a jet-sheath
configuration). Outside the light cylinder, the jet material propagates
outward super-Alfv\'{e}nically and cannot communicate backward with
Alfv\'{e}n waves. Because of this, above the initial oscillatory
phase, it behaves as part of an infinite, time-independent cylindrical
jet.
A snapshot of the cylindrical jet is shown in 
Figure ~\ref{fig:2D_headless_panels},  where we show from left to right,
meridional slices of
$b_\phi$, $b_p$, $b_\phi/b_p$ and $\gamma\beta$, respectively (the
colour schemes show $\log_{10}$ of the plotted parameters). Black
solid lines track equal poloidal magnetic flux surfaces and follow the
poloidal field lines.  The field lines, which stretch out from the central
compact object, make up the jet. They turn around at the tip of the jet
(which is located outside of Figure~\ref{fig:2D_headless_panels} frame)
and return back in a layer that surrounds the jet.  The returning
field lines carry no energy flux and they make up an {\it inner
  magnetised cocoon} that separates the magnetic jet from the funnel
wall. The jet and the inner cocoon are separated by a surface at
which the poloidal field lines switch direction and $b_p$ has a local
minimum, leading to a peak in the value of $b_\phi/b_p$. This surface
marks the jet outer boundary and is located in
Figure~\ref{fig:2D_headless_panels}(c) at $R_{\rm j}\simeq15$. The ratio
$b_\phi/b_p$ is close to unity in the jet and the Lorentz factor
increases from the jet axis toward the jet edge. Figure
\ref{fig:2D_profiles_headless} shows the profiles of $b_\phi$, $b_p$ and
$\gamma\beta$ across the jet at the height $z=200$. The strong $b_p$ 
with a comparable strength to $b_\phi$ outside the light cylinder ($R>1$) is clearly seen. Also evident is the flat profile of both field components, in agreement with eqn. \eqref{eq:b_phi_inf}, \eqref{eq:b_p_inf}, and the linear growth of  of $\gamma\beta$ which is expected in such a case (eq. \ref{eq:g_j_headless}). The combination of a flat magnetic field profile with a dominant poloidal component, together with the high Lorentz factors renders headless jets stable to kink instability.

\section{Axially symmetric headed jets: jets moving through a dense medium}\label{sec:2D}

Magnetised jets propagating in a dense stellar envelope need to drill
a hole through the star before they can emerge from it. Such
jets are substantially different than the \emph{headless}
jets that propagate in a previously evacuated funnel and that we
discussed in Sec.~\ref{sec:ideal}. When a jet propagates through a
dense external medium at a speed that exceeds the ambient sound speed,
it pushes the ambient gas in front of it,
forming a bow shock, as illustrated in  
Fig.~\ref{fig:headed_vs_headless}(a) with the solid blue line. The ambient gas that crosses the shock heats
up and forms a cocoon around the jet (shown in green in
Fig.~\ref{fig:headed_vs_headless}). The cocoon applies pressure on the jet
and collimates it, thus playing the role of the confining walls in the
previous example. In this case, however the jet continuously expends
energy on drilling the hole and pushing the ambient gas sideways.
This causes the jet to slow down mainly at its upper part, or the
\emph{jet head} (shown in dark blue in Fig.~\ref{fig:headed_vs_headless}), where most of the work is being made. 

\begin{figure}
\begin{center}
\includegraphics[width=\columnwidth]{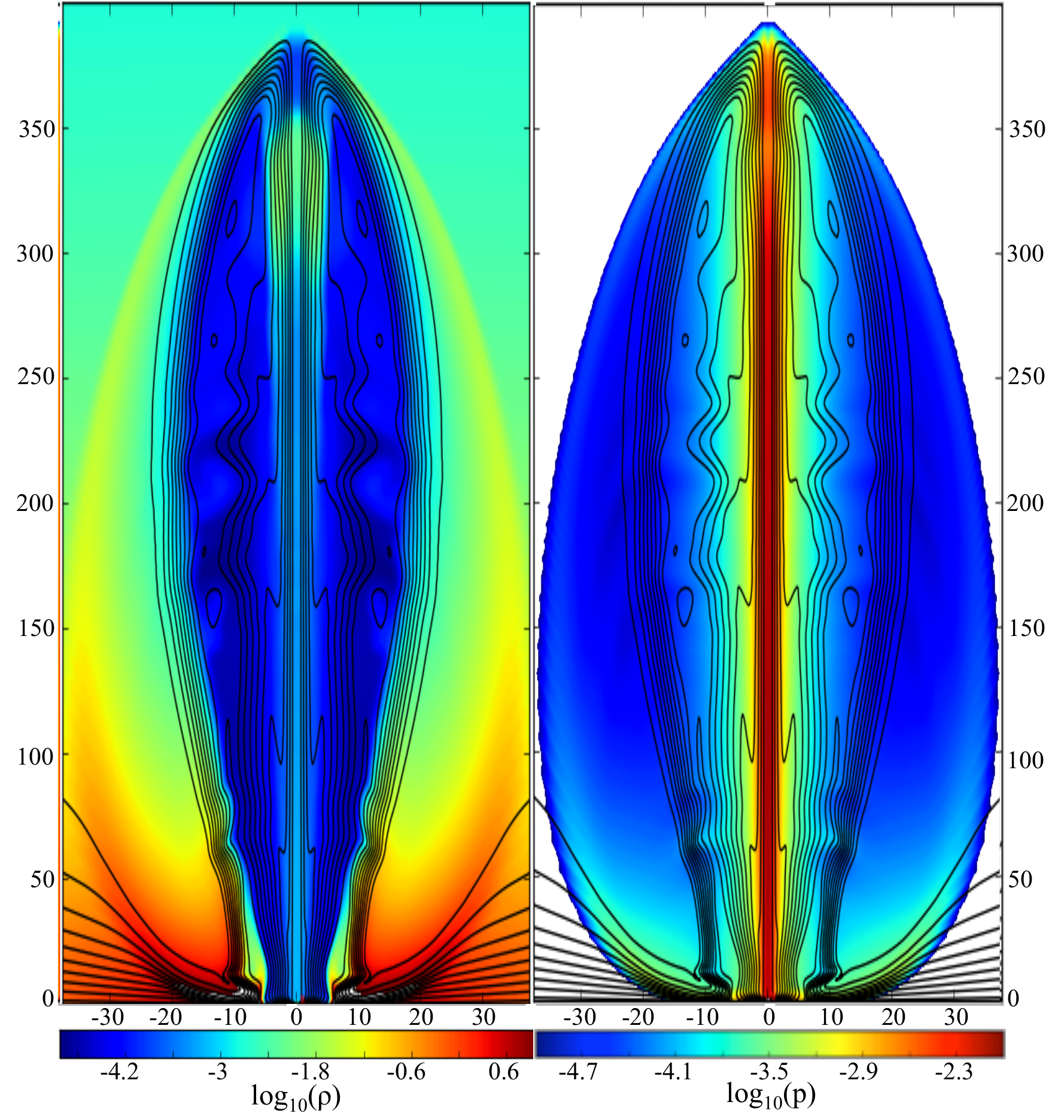} 
\end{center} 
\caption{Meridional slices through a 2D \emph{headed} jet, i.e., a
  jet that propagates through a dense ambient medium (without any
  pre-drilled funnel), at a time $t=730R_{\rm L}/c\sim0.25$~s in the fiducial 2D model M2. Density
  (left panel) and pressure (right panel) distributions are shown in
  colour on a log scale (see the colour bars). Solid black lines show magnetic field
  lines, which extend from the surface of the central compact object
  out to the jet head and return back on the outside of the
  jet. Because the jet head moves supersonically relative to the
  ambient sound speed, an arrowhead-shaped bow shock forms and appears
  as a sharp jump in colour in both panels. The ambient gas, which is
  heated at the shock, forms a \emph{cocoon} that pressure supports
  the jet and causes it to assume a cigar-like shape. The bow shock expands sideways with time, and the cocoon pressure decreases. As a result the jet slowly widens and never reaches a true steady state (compare
  to headless jets in Fig.~\ref{fig:2D_headless_panels}).}
\label{fig:2D_p_rho_jets}
\end{figure}

The basic features of the jet, the bow shock and the cocoon are seen
in Fig.~\ref{fig:2D_p_rho_jets} that shows a meridional slice of the
density (left panel) and the pressure (right panel) in a 2D simulation
of the jet. The shock is seen as a sharp jump in colour in the
density (left) and pressure (right) panels. To simulate the jet we
used a similar setup as in the headless jet case: a rotating sphere
with a monopole field that is placed at the center of the grid and is
surrounded by a cold ambient medium. However here the medium
completely surrounds the sphere and follows a power-law density (see
Table~\ref{tab:models_grid}, model M2). As in the previous case the
presence of the ambient medium causes the poloidal field lines to
collimate along the rotational axis. The poloidal field lines, shown
with black lines, extend from the jet base at the central object out
to the jet head, at which they turn around and return back. Similar to
the headless jet case, they form an inner magnetised cocoon that
shields the magnetic jet from the non-magnetic cocoon of the shocked
ambient gas.  The outgoing magnetic flux, which makes up the jet, has
an elongated, nearly cylindrical `cigar' shape: its radius at the
widest point is $R_{\rm j}\simeq 30$. The velocity of the jet head,
$\beta_{\rm h}\simeq 0.5$, is substantially lower than the velocity of the
jet material which is practically $c$, at the time shown in
Fig.~\ref{fig:2D_p_rho_jets}, as we discuss below.

To support the propagation of the jet head, the pressure in the jet
increases, mainly due to the increase in the toroidal field strength.
Another way to look at this is that the slower moving head blocks the
free expansion of toroidal field which otherwise would stream upward
with a super-fast magnetosonic velocity obtained in
eq.~\eqref{eq:g_j_headless}. Since toroidal field keeps getting injected into the
jet by the winding of the field lines at the bottom of the jet, the
toroidal field accumulates in the jet, and the ratio of toroidal to
poloidal field increases. %

The jet and the cocoon are in pressure balance, and the width of the
jet is determined by the pressure profile of the cocoon along the
jet-cocoon boundary. This profile is relatively flat in the
longitudinal direction, leading to the near-cylindrical
cigar-like jet shape seen in
Fig.~\ref{fig:2D_p_rho_jets}. Moreover, the cocoon expands sideways
with a velocity that is much slower than the velocity of the jet
material along the jet. Thus, the typical timescale for the cocoon pressure
to drop by a factor of two is much longer than the timescale for a jet
fluid element to double its height. This implies that instantaneously
we can approximate the jet geometry as a steady-state cylindrical jet,
and we can use the force-balance equation \eqref{eq:dbdr} to analyse
the transverse profile of the magnetic field in the jet. 

Here again we need to know the relationship between $b_\phi$ and
$b_p$.  Due to the accumulation of toroidal field in the jet, we approximate $b_\phi\gg b_p$ at
$R\gg R_{\rm L}$, which leads to the following solution to eq.~\eqref{eq:dbdr}:
\begin{equation}\label{eq:b_phi_prop}
  b_\phi\propto
  \left\{
\begin{array}{ll}
 R & \mbox{, $R\ll R_{\rm L}$ },\\
 R^{-1} & \mbox{, $R\gg R_{\rm L}$ }.
\end{array}
\right.
\end{equation}
The excess of toroidal over poloidal field implies that the acceleration of the plasma cannot be very efficient. Substituting the approximation $b_\phi\gg b_p$ in eq. \eqref{eq:gamma_d}, we get that the drift 4-velocity is now
\begin{equation}
\gamma_{\rm j}\beta_{\rm j}\approx \frac{E}{(B_\phi^2-E^2)^{1/2}}\simeq 1, \quad \text{(headed jets)}
\label{eq:gammabeta_headed}
\end{equation}
where we assume here that $|B_\phi|$ exceeds $E$ by an order unity factor due to the accumulation of toroidal field.
In this case eq.~\eqref{eq:Bphi_infinite} which is still correct in the lab frame, holds approximately for the fluid-frame as well, namely $b_p\simeq b_\phi R_{\rm L}/R$.
This implies that
\begin{equation}\label{eq:b_p_prop}
  b_p\propto
  \left\{
\begin{array}{ll}
 \mbox{constant} & \mbox{, $R\ll R_{\rm L}$ },\\
 R^{-2} & \mbox{, $R\gg R_{\rm L}$ }.
\end{array}
\right.
\end{equation}
Comparison with eq.~\eqref{eq:b_phi_inf} shows that the profiles of
both components of the magnetic field in this case are steeper at
$R>R_{\rm L}$ than in the case of a headless jet. 
 
\begin{figure}
\begin{center}
\includegraphics[width=\columnwidth]{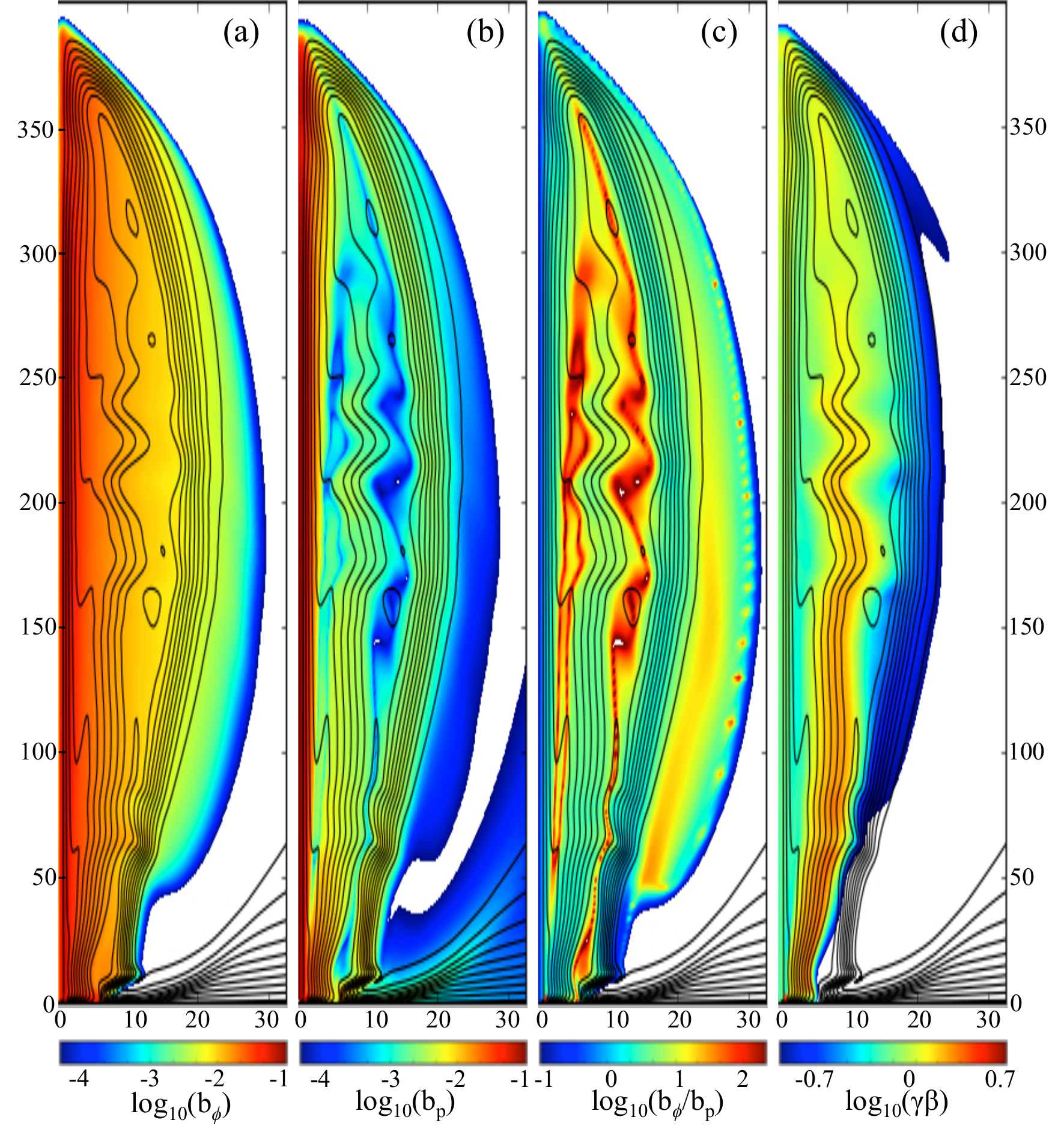}
\end{center}
\caption{The 4 panels show, from left to right, meridional slices of
    $b_\phi$, $b_p$, $b_{\phi}/b_p$ and $\gamma\beta$ of
    a headed jet drilling its way through a dense ambient medium, at a
    time $t=730R_{\rm L}/c\sim0.25$~s in our fiducial 2D model M2. The
    colour scheme is the same as for the headless jet in
    Fig.~\ref{fig:2D_headless_panels} for ease of comparison. Poloidal field lines are shown as
    black solid lines. The edge of the jet, the boundary between
    outgoing and returning field lines, is located at
    $R_{\rm j}\simeq10{-}15R_{\rm L}$. Along the jet edge the poloidal field reaches
    a minimum and the ratio $b_\phi/b_p$ reaches a maximum (panel c). See
    Fig.~\ref{fig:2D_profiles_headed} for profiles of various
    quantities across the jet.}
\label{fig:2D_grb_jets}
\end{figure}

To illustrate the magnetic field properties and the dynamics in the headed jets, we show in Fig.~\ref{fig:2D_grb_jets}(a)--(d) the spatial distributions of $b_\phi$, $b_p$, $b_\phi/b_p$, and $\gamma\beta$ of the jet shown in
Fig.~\ref{fig:2D_p_rho_jets}. The colour bars are the same as in
Fig.~\ref{fig:2D_headless_panels}, to allow for easy comparison to the
headless jets. Panel by panel comparison of
Fig.~\ref{fig:2D_grb_jets} and Fig.~\ref{fig:2D_headless_panels} shows
that in headed jets $b_\phi$ is stronger (panel a) and $b_p$ is much more
concentrated toward the axis (panel b) in the headed jet, as expected
from our analytic scalings of the field profiles (see
eqs.~\ref{eq:b_phi_inf}--\ref{eq:b_p_prop}). Therefore $b_\phi/b_p$ (panel c) is much higher in the headed jet than in the headless jet. The two red
stripes that appear at $R\simeq5$ and $R\simeq15$ mark regions where
the poloidal field flips direction. 
The 4-velocity (panel d)
is much lower in the headed jet than in the headless jet, as expected due
to the need for the headed jet to drill through the ambient medium.

\begin{figure}
\begin{center}  
    \includegraphics[width=\columnwidth]{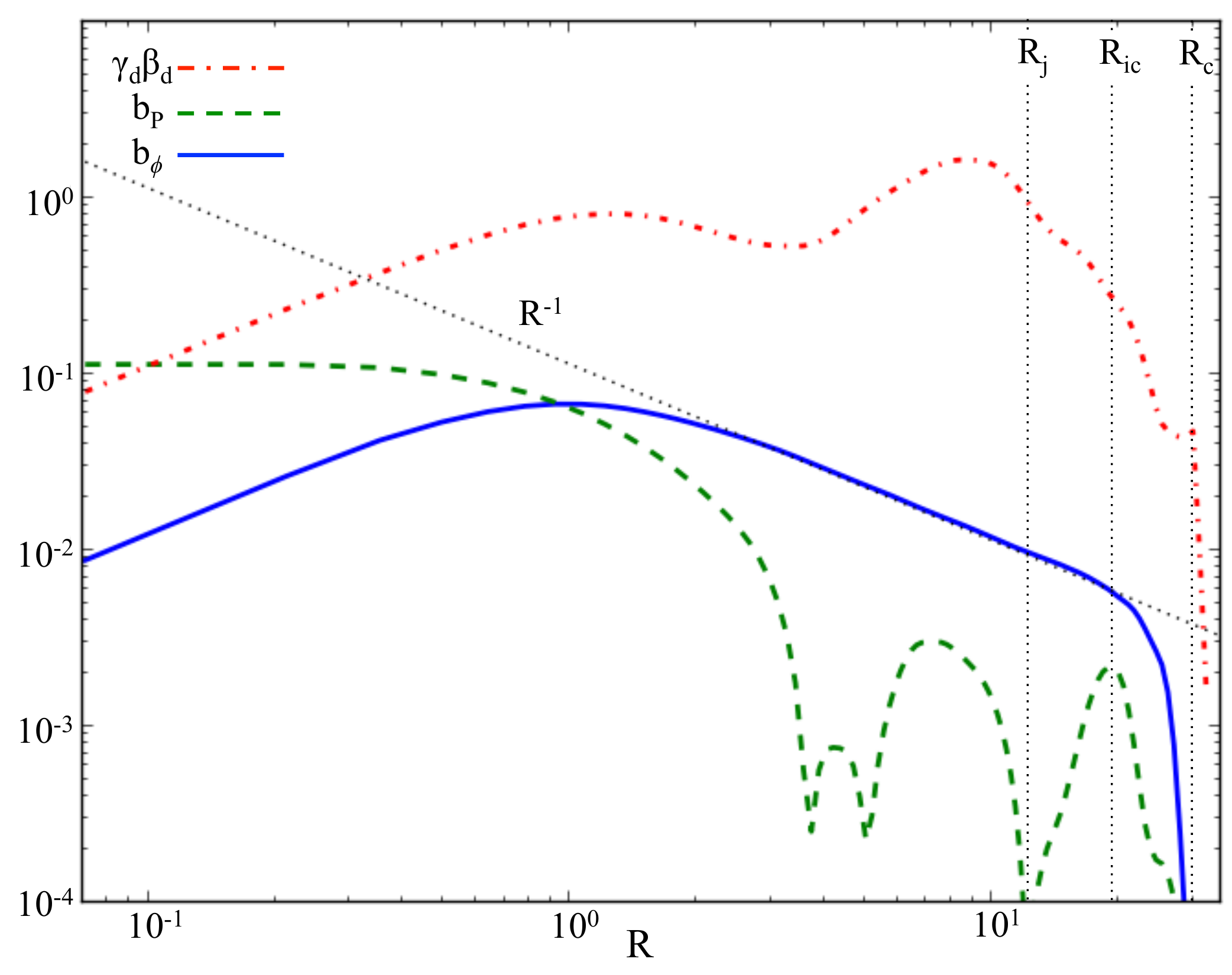}
\end{center}
\caption{The  cross-sectional profiles of the drift 4-velocity (dash-dotted red)
  and the two fluid-frame magnetic field components,
  $b_p$ (dashed green)  and $b_\phi$ (solid blue), in a headed 2D jet (model M2) at an altitude $\abs{z}=200R_{\rm L}\sim2\times10^{9}$~cm.
  Here, the fluid-frame magnetic field scales as $b_{\phi}\propto R^{-1}$
  and peaks around the light cylinder, $R_{\rm L}=1$. The profile of
  $b_p$ is flat inside and drops sharply at $R \gg R_{\rm L}$, where the
  toroidal field dominates, $b_\phi\gg b_p$. As a result, the drift velocity
  $\gamma_{\rm d}\beta_{\rm d}$ remains of order unity across most of the
  jet. This is in
  contrast to the headless jet that accelerates to much higher
  velocities (see Fig.~\ref{fig:2D_profiles_headless}). The radii of
  the jet, which contains the outgoing magnetic field lines ($R_{\rm j}$), the inner
  cocoon, which contains the returning field lines ($R_{\rm ic}$), and the outer
  cocoon, which is made up of shocked ambient gas
  ($R_{\rm c}$), are marked with dotted vertical lines. The full morphology of the headed jets is shown in
  Figs.~\ref{fig:2D_p_rho_jets}--\ref{fig:2D_grb_jets}.}
\label{fig:2D_profiles_headed}
\end{figure}

Figure~\ref{fig:2D_profiles_headed} shows the jet properties in a more quantitative way by plotting the 1D profiles of $b_p$,
$b_\phi$ and the drift 4-velocity $\gamma_{\rm d}\beta_{\rm d}$ across the jet.
The profiles are measured at $z=200R_{\rm L}\sim2\times10^{9}$~cm, the widest point of the
headed jet where it has a similar width to the headless jet and a
geometry that is close to cylindrical, making it easy to compare the
two cases (see
Figs.~\ref{fig:2D_p_rho_jets}--\ref{fig:2D_grb_jets}). It can be seen
that $b_\phi$ peaks at $R\sim R_{\rm L}$ and scales as $b_\phi \propto R^{-1}$ beyond that, in agreement with eq.~\eqref{eq:b_phi_prop}. This profile is steeper than for the headless
jet, $b_\phi \propto R^{-1/2}$, as seen in
Fig.~\ref{fig:2D_profiles_headless}. On the other hand, $b_p$ has a
flat core at $R<R_{\rm L}$ and drops sharply outside of it to a very low value,
$b_p\ll b_\phi$, unlike in the headless case for which
$b_p\simeq b_\phi$. As a result, the acceleration is much less
efficient in headed jets, with $\gamma_{\rm d}\beta_{\rm d} \lesssim 2$ across the jet.
 The combination of these three properties -- higher $b_\phi/b_p$
ratio, steep profile of $b_p$, and lower $\gamma\beta$, suggests that
headed jets are unstable to kink modes. As we show in the next section, this leads to jets that are very different than the headless jets.

\section{Kink Modes in Headed Poynting-dominated
  Jets}
\label{sec:3D}

Kink instability is a 3D effect and cannot be studied with
axially-symmetric simulations. It develops in regions that
are dominated by the toroidal field (in the proper frame) and that
maintain strong causal connection accross the jet
(eq. \ref{eq:gammatheta}).  
For this reason, collimated jets are ideal environments for the development
of the kink instability. Here we show that the instability
evolves in two stages.  First, it grows internally in the
jet without affecting the overall jet morphology. This instability is
often referred to as the {\it internal kink}.  It converts the magnetic energy into thermal
energy via magnetic reconnection. As a result, the toroidal magnetic field
decays, and the jet finds itself in a stable configuration that inhibits 
further growth of the internal kink. However, kink modes can still
grow externally on the periphery of the jet and perturb the entire jet
body. Such an {\it external kink} instability grows over longer time
scales and typically affects the outer parts of the jet, near the
jet head. As we will see below, as a result of the external kink, the jet head
wobbles and increases its effective cross-section. This
makes it harder for the jet to drill its way through the ambient
medium and decreases the jet propagation velocity.

\subsection{The growth of internal kink in a collimated headed jet}\label{ssec:internal_kink}

An outflow from the central compact object is initially over-pressured relative to the surrounding medium and expands in all directions. As the flow expands, its pressure drops until it becomes comparable to the surrounding pressure of  the confining medium, and the outflow becomes collimated.

Before the collimation point, the jet internal pressure exceeds the pressure of the cocoon, so the jet material expands freely.  The Lorentz factor scales as
$\gamma_{\rm j}\propto R/R_{\rm L}$ (eq. \ref{eq:g_j_headless}) until
$\gamma_{\rm j}=\theta^{-1}$. From this point on, the jet material loses the
strong causal contact with the jet axis and the acceleration continues
less efficiently \citep{tch08}.  During the free expansion phase kink modes cannot grow, since their typical growth time is longer than the propagation
time. Indeed, from eq.~\eqref{eq:t_kink}, it follows that in the
linear acceleration regime $t_{\rm kink}\approx 2\pi R^2/R_{\rm L}c > R/c
\sim t$. Beyond that the loss of strong causal contact inhibits the
growth of the kink instability.
 
At the collimation point, the jet pressure becomes equal to the
pressure in the cocoon.  The conditions for strong causal contact are remet and the kink instability can grow.
If the collimation takes place outside of the linear acceleration
regime, then at the collimation point we have $b_\phi > b_p$. The hoop
stress, which becomes effective and is proportional to $b_\phi^2$ (see
eq. \ref{eq:dbdr}), cannot be counterbalanced by the poloidal
pressure. This leads to a recollimation point---contraction of the jet cross section---followed by a bounce when the poloidal pressure becomes strong enough to resist the hoop stress, resulting in a nozzle--like shape \citep{lyub09}.  
In a headless jet its cylindrical radius continues to oscillate above the nozzle until 
the magnetic field configuration relaxes to
an equilibrium state, described by eq.~\eqref{eq:b_phi_b_p_inf}, and seen in Fig. \ref{fig:2D_headless_panels}. Since $b_\phi\simeq b_p$, this state is
stable against internal kink modes \citep{lyub09,2011PhRvE..83a6302L}.
In a headed jet, as Fig.~\ref{fig:2D_grb_jets} shows, these two
components are not in equilibrium
with each other, and the
magnetic field continues to be dominated by $b_\phi$
above the collimation point (at $R\geq R_{\rm L}$). This
suggests that the jet can be unstable to the internal kink instability.

\begin{figure}
  \begin{center}
\includegraphics[width=\columnwidth]{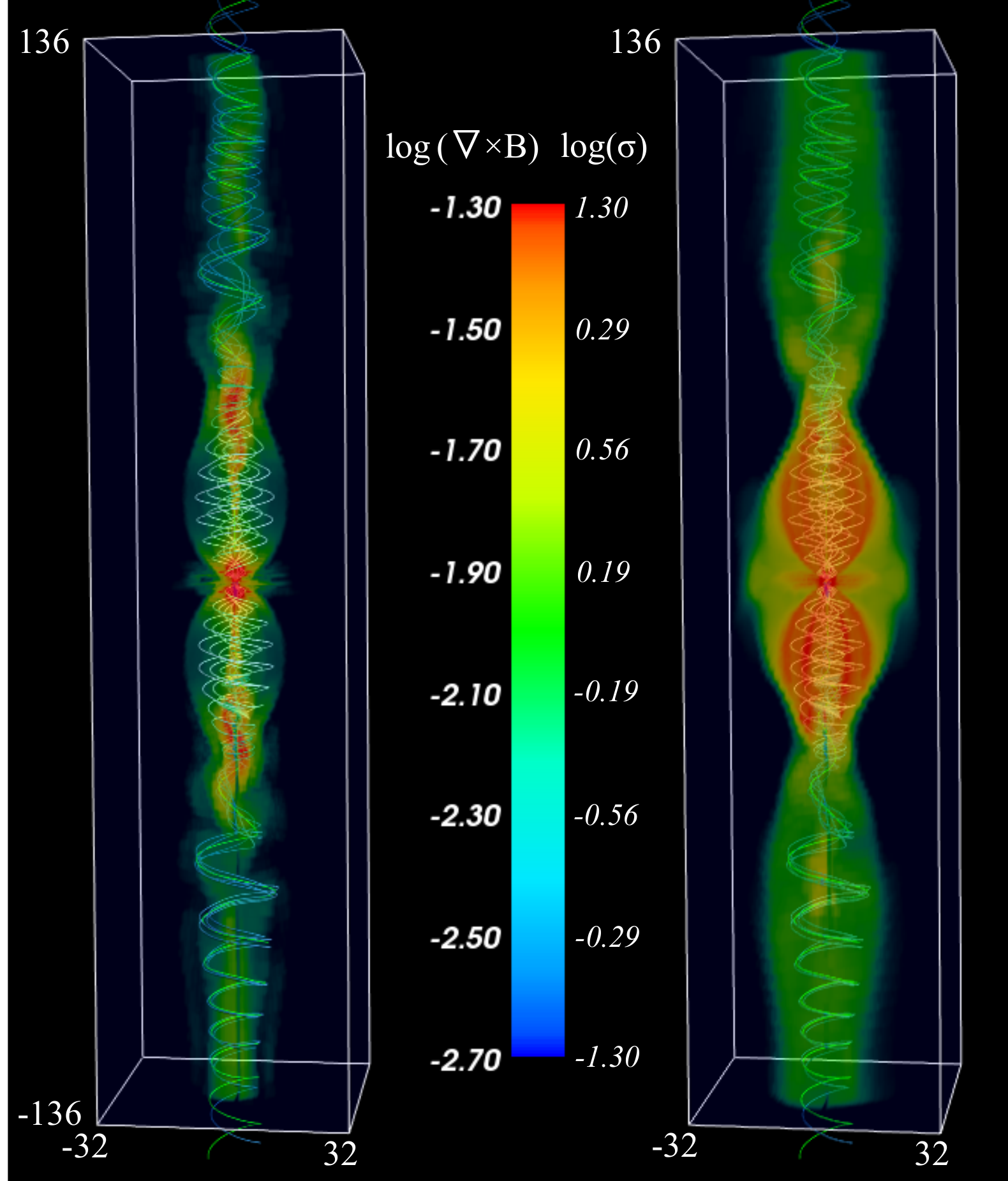}
\end{center}
\caption{A snapshot of the central region in our fiducial 3D model M3 at
  $t=4400R_{\rm L}/c\sim1.5$~s, when the jet head is at
  $z=1000R_{\rm L}\sim10^{10}$~cm, or about $10\%$ of the stellar radius. The colour scheme in the left panel shows the
  $\log_{10}(\nabla\times\myvec{B})$, which is a tracer of conduction
  currents, and the right panel shows the $\log_{10}(\sigma)$, which is
  a tracer of magnetisation. The solid lines how the field lines of
  $\myvec{b}$, the magnetic field \emph{as measured in the fluid
    frame} and traced out \emph{in the lab frame}. The degree of their
  twist reflects the degree to which the toroidal
  magnetic field component dominates in the fluid frame. The
    jet is initially freely expanding sideways out to    $\left|z\right|\simeq 30R_{\rm L}$
    where it recollimates due to the confining pressure of the
    cocoon. The free
  expansion region is characterised by a low
  current density (green colour) and a high $\sigma$ (red colour; see
  the colour bar).  Above this region, $\sigma$ approaches unity
  (green colour) and the field lines become less toroidally twisted.  Both of
  these reflect the dissipation of the toroidal magnetic field into
  heat above the collimation point. There the magnetic field lines
 converge to the axis and become unstable to the internal kink mode,
  which seen as a helical pattern in the jet (see
  Sec. \ref{ssec:internal_kink}). }
\label{fig:3D_jet_fieldline}
\end{figure}

To check this we carried out 3D simulations whose setup is similar to
that of our 2D simulations, with a few exceptions, as described in
Sec.~\ref{sec:3d-simulation-setup}.\footnote{In 3D simulations, the
  rotational axis is directed not along the $z-$axis, as in 2D ones,
  but along the $x-$axis, to avoid the interaction of the jets with
  the coordinate singularity. Accordingly, the radial 3D grid lines
  collimate toward the $x-$axis. For simplicity of presentation, in
  the text we will still use the standard axis orientation, with the
  $z-$axis pointing along the rotational axis.} For our fiducial model, M3, we use an ambient density that corresponds to a GRB jet with a luminosity $5\times10^{49}$ erg/s propagating in a $10M_\odot$ star (see Tabs.~\ref{tab:models_grid} and \ref{tab:models_properties} for a full list of models). 
Figure \ref{fig:3D_jet_fieldline} shows a volume rendering of the inner
regions of a headed jet in our fiducial model M3.
The left panel shows in colour the quantity $\log_{10}(|\nabla\times\myvec{B}|)$, which is a proxy for
the conduction current density.  The right panel shows the logarithm
of magnetisation, defined here as $\sigma=B^2/4\pi w$. The light blue
lines trace out the magnetic field in the lab frame ({\rm\bf B}). 
Close to their footpoints, at $z\approx0$, the field lines expand and
have an ordered, toroidally-dominated configuration. The field lines
are recollimated at $z\simeq20R_{\rm L}\simeq 2\times10^8$~cm and converge to the jet axis and
rebound, forming a nozzle-like shape. As they approach
the nozzle, they begin to kink and lose their ordered shape. 
This is accompanied with a drop in the $\sigma$ parameter and a peak
in the conduction current density, indicative of magnetic
dissipation. The helical pattern of the dissipation region is
characteristic of the kink instability. Above the dissipation region
the field lines are less twisted and have a stronger poloidal
component, which points to a dissipation that affects mostly the
toroidal magnetic field component.

Not all field lines lose the strong causal contact with the axis in
the free expansion region.  If the jet is collimated at an altitude
$z_{\rm coll}$ then strong causality is maintained across the field lines
with an opening angle $\theta<\sqrt{R_{\rm
    L}/z_{\rm coll}}$ %
up to $z_{\rm coll}$. If, in addition, the Lorentz factor along the field
lines is not too high, $\gamma_{\rm j}<\sigma_0^{1/3}$, the flow remains
sub-fast magnetosonic. In this case it can ``feel'' the decelerating material at the recollimation nozzle, above it.
This
has a similar effect as the slower moving head and leads to an
increase in the toroidal field strength: toroidal field accumulates,
the acceleration becomes inefficient, and the region becomes kink
unstable. The two conditions are met along field lines with an opening
angle satisfying
\begin{equation}\label{eq:theta_strong}
  \theta < \theta_{k}=
\left\{
\begin{array}{ll}
 \sqrt{\frac{R_{\rm L}}{z_{\rm coll}}} & \mbox{, $z_{\rm coll}<R_{\rm L}\sigma_0^{2/3}$},\\
 \\
 \frac{R_{\rm L}}{z_{\rm coll}}\sigma_0^{1/3} & \mbox{, $z_{\rm coll}\geq R_{\rm L}\sigma_0^{2/3}$},  \ 
\end{array}
\right.
\end{equation}
and render the field lines kink
unstable %
below the recollimation nozzle. 

\begin{figure} \begin{center}
    \includegraphics[width=\columnwidth]{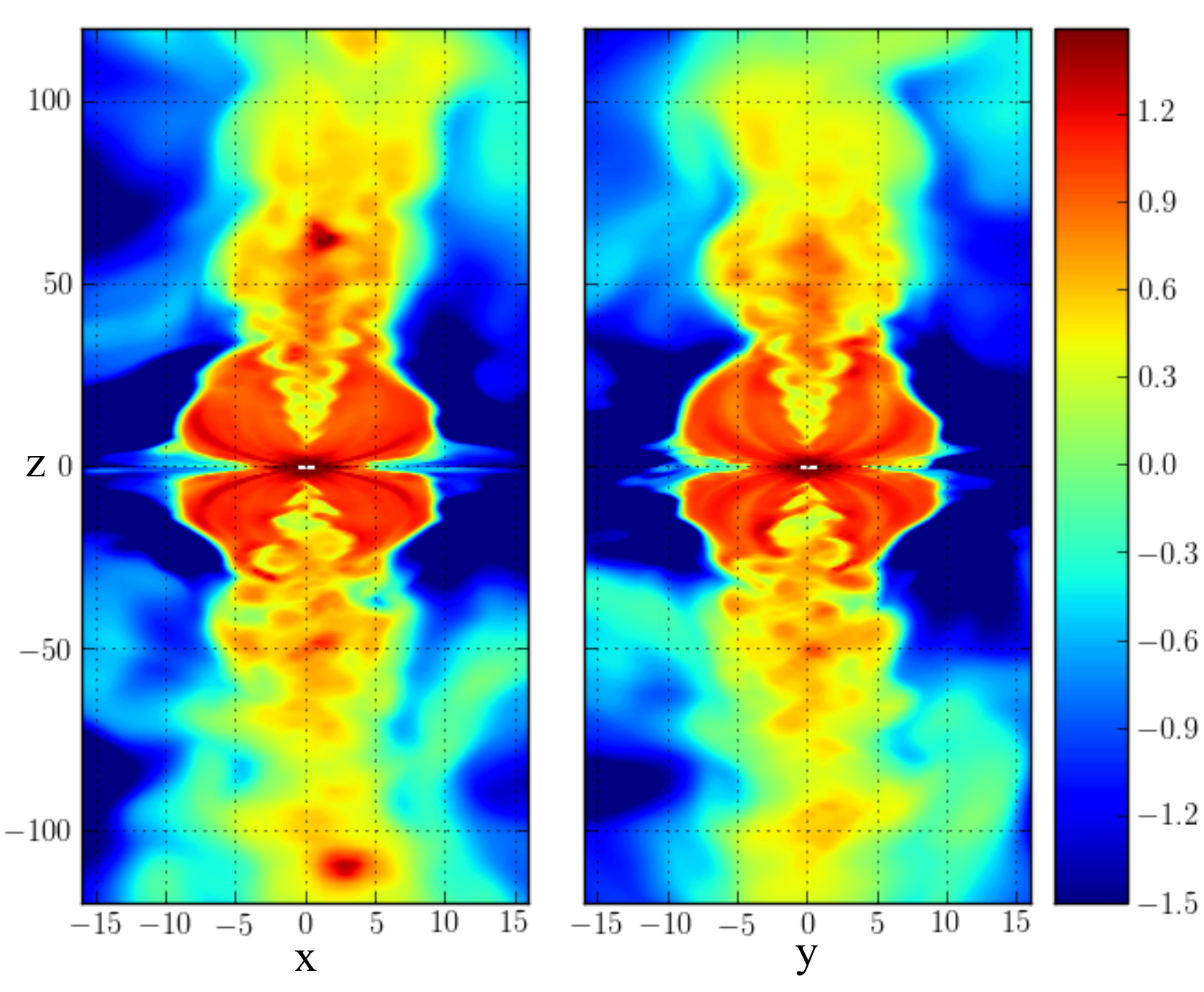}
  \end{center} \caption{Meridional slices of $\log_{10}(\sigma)$
    through the $x-z$ (left) and $y-z$ (right) planes, in a 3D jet
    at time $t=2984R_{\rm L}/c\approx 1$~s in our high-resolution 3D simulation, model
    M3HR. Note that the horizontal axis scale has been stretched by
    approximately an order of magnitude to clearly show the jet
    structure. The jets start out highly
    magnetised, with $\sigma_0\simeq25$. They recollimate at
    $z_{\rm coll}\simeq 17R_{\rm L}\approx1.7\times10^8$~cm, converge onto the axis, and develop an
    internal kink instability, which is seen as small-scale yellow-red
    wiggles in the jets. The decrease of the magnetisation (seen as the
    transition in colour from red to yellow) at the
    wiggles reflects magnetic energy 
    dissipation via the internal kink instability. Some field lines,
    which have small enough opening angles to maintain
    strong causal contact across the jets,
    become unstable earlier and dissipate their energy at lower
    altitudes. These lines form a `wedge' of lower $\sigma$ that
    extends down to $\abs{z}\simeq10R_{\rm L}\approx10^8$~cm.}
    \label{fig:sigma_high}
\end{figure}

\begin{figure}
\begin{center}
\includegraphics[width=\columnwidth]{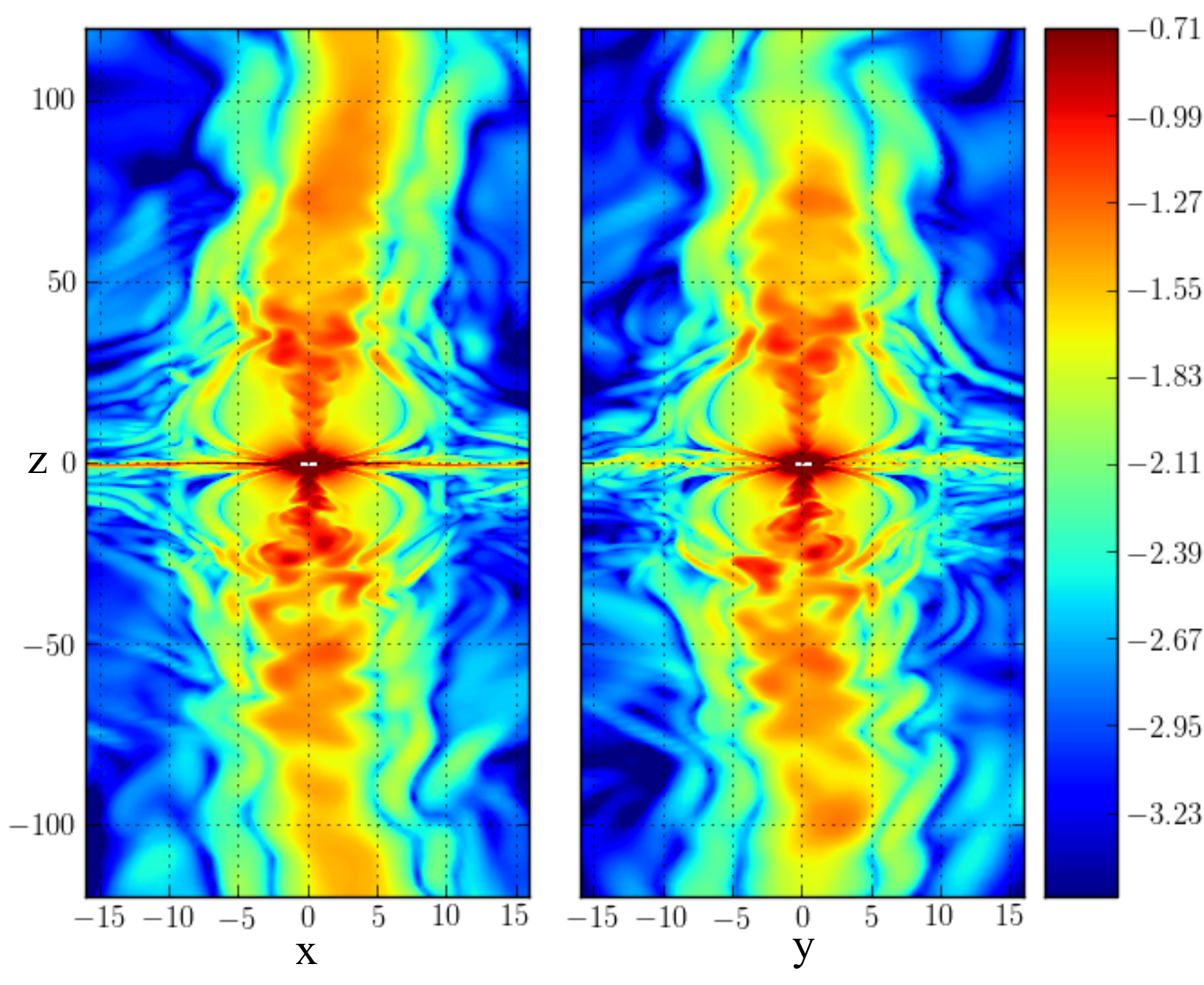} 
\end{center}
\caption{Same as Fig.~\ref{fig:sigma_high} but for meridional slices
  of the conduction current $J \propto \log_{10}\left(|\nabla\times
    B|\right)$. High (red) values of
  $J$ indicate regions with particularly strong magnetic field twist
  and particularly strong dissipation.}
    \label{fig:J_high}
\end{figure}

To study this effect in more detail and to better capture the
small-scale structure of the dissipation region, we reran our fiducial 3D model M3
at twice the resolution in each dimension. We refer to this
high-resolution simulation as model M3HR  (see Table~\ref{tab:models_grid}).
Figures \ref{fig:sigma_high}
and \ref{fig:J_high},  show the longitudinal cross-sections of the jet along the $y{-}z$ plane (left panel) and the $x{-}z$ plane (right
panel) in this high-resolution model. The jet propagates along the $z-$axis. The colour schemes are
the same as in the corresponding 3D jets in
Fig.~\ref{fig:3D_jet_fieldline}. Regions where magnetic energy is
dissipated are identified by a decrease in the $\sigma$ parameter
accompanied by the high conduction current. Below the nozzle there is
a cone of an opening angle $0.1\simless\theta\simless0.17$, which is kink unstable
all the way down to the source. The jet becomes collimated at
$z_{\rm coll}\simeq17$ (see Fig.~\ref{fig:J_high}), the opening angle of
the unstable cone is in agreement with eq.~\eqref{eq:theta_strong} for
our choice of the initial jet magnetisation of $\sigma_0=25$ (see Sec.~\ref{sec:scheme}).

\begin{figure}
  \begin{center}
\includegraphics[width=\columnwidth]{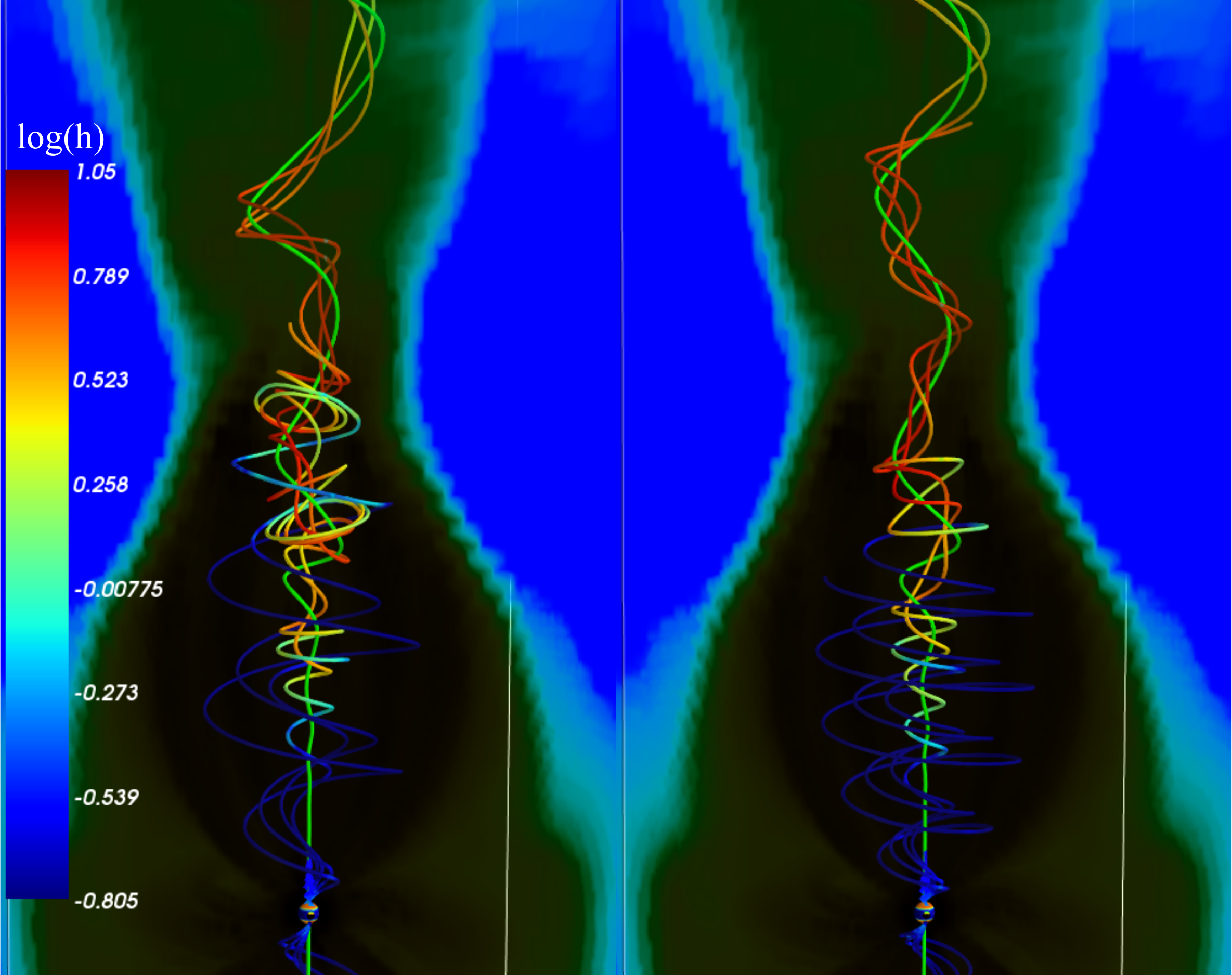} 
\end{center}
\caption{A zoom-in on the jet recollimation nozzle that is seen in
  Fig.~\ref{fig:3D_jet_fieldline} at $z\sim50R_{\rm L}\simeq 5\times10^8$~cm (model M3 at
  $t = 4400R_{\rm L}/c \sim 1.5$~s). The left panel shows field lines in the
  fluid-frame $\myvec{b}$ as traced out in the lab frame. This
  unconventional representation of the magnetic field lines is useful
  for visually determining the degree of toroidal dominance of the
  magnetic field in the fluid-frame (see the text for details). The
  right panel shows the conventional field lines of the lab-fame
  magnetic field, $\myvec{B}$. The fluid-frame magnetic field lines
  are more disordered, which might reflect their readiness to
  reconnect and dissipate their energy. We show the field lines that originate
  at the surface of the magnetar with polar angles at their foot
  points equal to $0^\circ$, $20^\circ$, $30^\circ$, $50^\circ$, and
  $70^\circ$.
  The colour scheme represents $\log_{10}h$ on the field lines,
  where $h$ is the specific enthalpy. The volume colour rendering
  represents $\log_{10}\sigma$. The jet has a higher $\sigma$ than
  the confining cocoon and appears as a dark silhouette against
  the blue background. The field line, which originates at the north
  pole of the magnetar, indicates the jet axis and is shown in
  green. The gradual straightening of the magnetic field lines along
  the green line is clearly seen in both panels and reflects the
  dissipation of the toroidal magnetic field due to the internal kink
  instability.}
\label{fig:3D_jet_fieldlines2}
\end{figure}

Most of the magnetic energy dissipation occurs at the collimation
nozzle and above it. In this region magnetic field lines with
different pitch angles converge and become kink unstable. 
The dissipation process that follows leads to the straightening of the field lines, by reconnecting out the local toroidal component and results in a gradual alignment of the field lines with the helical path of the jet axis.
An example for such small angle reconnection can be seen in
Figure~\ref{fig:3D_jet_fieldlines2}. It shows a closeup of the field
lines at the collimation point, where most of the reconnection
occurs. We trace out several magnetic field lines starting out with polar angles
of $\theta=0^\circ, 20^\circ, 30^\circ, 50^\circ, 70^\circ$ on the
rotating sphere. The left panel of Fig.~\ref{fig:3D_jet_fieldlines2} shows the field
lines in an unconventional sense: it shows the field lines of \emph{proper
frame magnetic field}, $\myvec{b}$, that are traced out in the lab frame.
These field lines are useful for determining the regions that are
prone to becoming kink-unstable: this is because the growth of the kink
instability depends on the ratio $b_\phi/b_p$, which is seen in the
figure as the degree of azimuthal winding of the field lines. These
field lines also give an idea of the proper angle between two
reconnecting field lines at the reconnection point. The right panel of
Fig.~\ref{fig:3D_jet_fieldlines2} shows the conventional lab-frame
field lines $\myvec{B}$. From the comparison of the two panels, we
see that the proper field lines appear to be more disordered than the lab frame
ones, and this can be a factor in encouraging their rapid reconnection.
The  volume rendering colour scheme shows the logarithm of the
magnetisation, $\log_{10}\sigma$.
The the high-$\sigma$ jet appears as a dark silhouette against the blue
low-$\sigma$ background. The colour  of the field lines
represents $\log_{10}h$, where $h=w/\rho$ is the specific
enthalpy. Regions where the field lines are turning red are the regions
where the dissipation is taking place.
The green line in each of the two panels indicates the polar field line, which starts at the north pole of the magnetar and follows the centre of the jet.  As expected from the causality condition~\eqref{eq:theta_strong} and seen in Fig.~\ref{fig:3D_jet_fieldlines2}, the green line
begins to kink even before the jet converges onto the jet axis.  The
field lines with larger opening angles begin to kink only after the
flow converges onto the axis. These outer field lines are
initially strongly twisted around the central, green field
line. However, as they dissipate their magnetic energy into heat, they gradually straighten out and become aligned with the green line (as seen in Fig.~\ref{fig:3D_jet_fieldlines2}). 

\subsection{External kink above the dissipation region}\label{ssec:equipart}

\begin{figure}
  \begin{center}
  \includegraphics[width=\columnwidth]{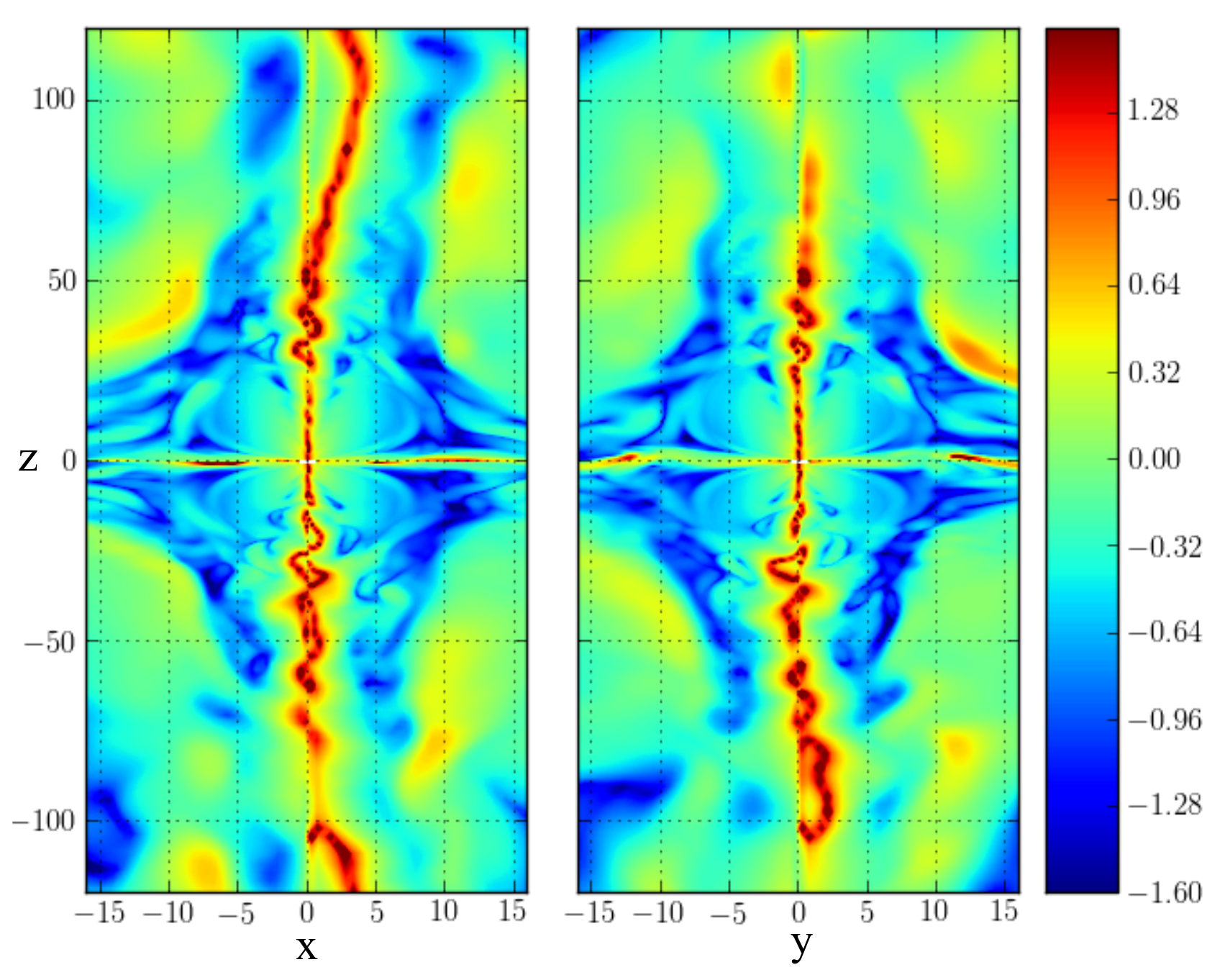}  
  \end{center}
  \caption{Similar to Figures~\ref{fig:sigma_high}
    and~\ref{fig:J_high} but for meridional slices of
    $\log_{10}(b_p/b_\phi)$. Dark red colour shows the
    poloidally-dominated regions and their transverse extent reflects
    the value of $R_0$, the cylindrical radius of the jet core with
    $b_p\gg b_\phi$.  We see that above the collimation point,
    $\left|z\right|\gtrsim 30R_{\rm L}$, the value of $R_0$ increases with increasing
    distance away from the central compact object. This increase
    reflects the conversion of the toroidal magnetic field into heat
    by the internal kink mode. Notice that the top jet in the right
    panel moves off the image plane due to the external kink.}
    \label{fig:b_ratio_high}
\end{figure}

The dissipation of magnetic field above the collimation point leads to two important
outcomes. First, it relieves some of the toroidal pressure and decreases the ratio of $b_\phi/b_p$ in the jet. As a result, $R_0$, the radius of the poloidal field dominated core, which controls the minimal growth time of the kink instability (see eq. \ref{eq:t_kink}), increases significantly. This process is depicted in
Fig.~\ref{fig:b_ratio_high} which shows $\log_{10}(\abs{b_p/b_\phi})$ in a
cross section cut along the jet: the core starts out
as a narrow spine at the base of the jet and gradually widens
above the collimation point, at which most of the dissipation
occurs. Second, the dissipation generates thermal energy at the
expense of toroidal magnetic field energy. In the presence of the thermal pressure, the transverse magnetic field profile flattens out.
As we discussed in Sec.~\ref{sec:Primer}, both of these changes
stabilise the jet against the further growth of kink modes at the jet
inlet. In fact, we see that the internal kink saturates and the
dissipation stops when the equipartition is reached between the
thermal pressure and the magnetic pressure. 

\begin{figure}
  \centering
\includegraphics[width=8.5cm]{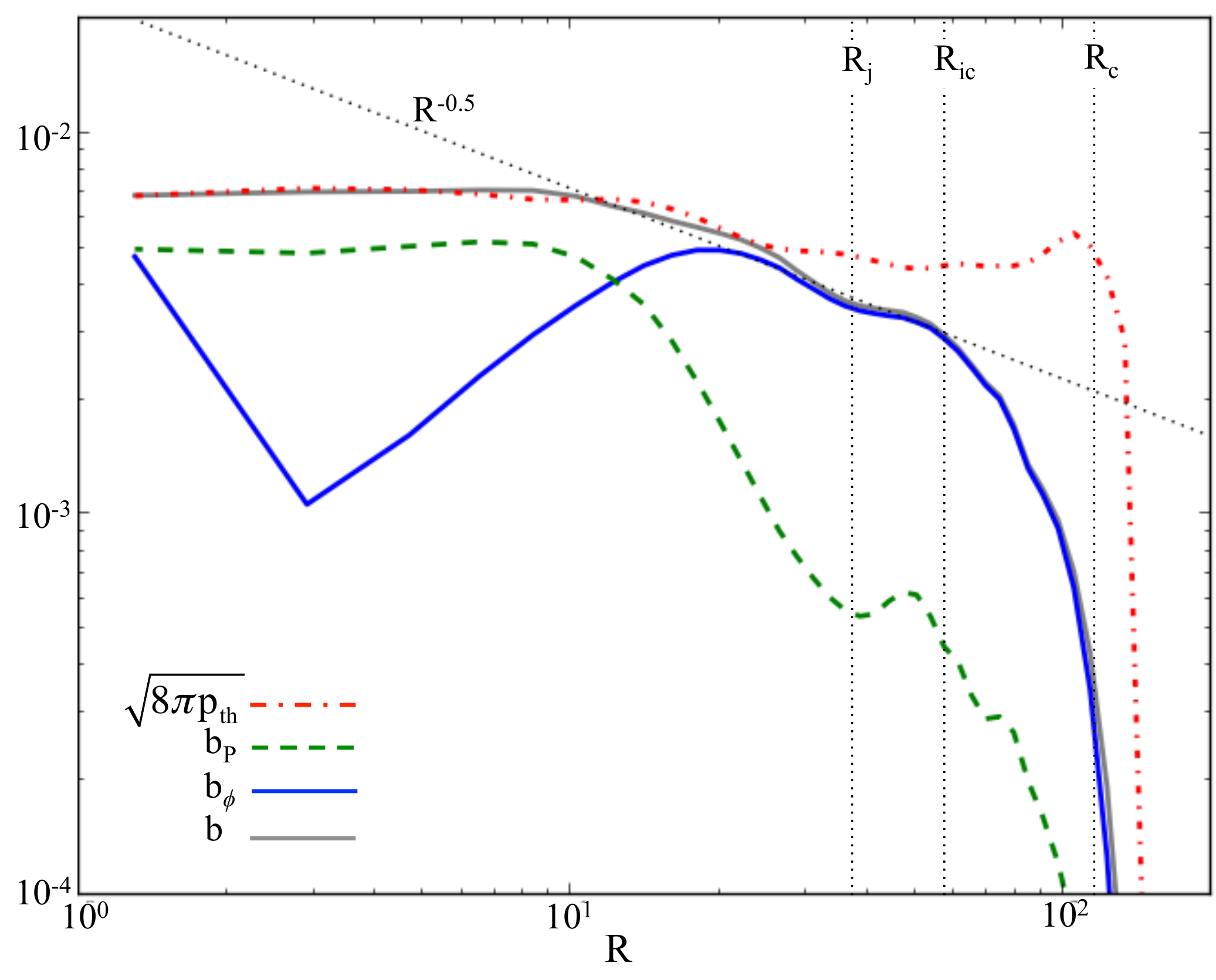}
\caption{ 
  Transverse profiles of the magnetic field components and the
  thermal pressure in the jet shown in
  Fig.~\ref{fig:3D_jet_fieldline} (fiducial model M3), taken at an altitude $z=400R_{\rm L}\sim
  4\times 19^9$~cm, $t
  = 4400R_{\rm L}/c\sim1.5$~s, and averaged over the
  azimuthal direction to smooth out non-axisymmetric effects.
  We show $\avg{b_\phi}$ (solid blue), $\avg{b_p}$ (dashed green), $\avg{b}=\avg{\sqrt{b_p^2+b_\phi^2}}$ (solid grey) and
  $\avg{\sqrt{8\pi p_{\rm th}}}$ (dash-dotted red). Here, the poloidal and
  toroidal field components are defined with respect to the magnetar's
  rotational axis, and the actual jet axis might deviate from it due
  to the external kink mode. The magnetic field
  profile consists of a core of radius $R_0\sim10R_{\rm L}\simeq 10^8$~cm, which is filled with a nearly uniform
  poloidal magnetic field, and is surrounded by a sheath, which is
  dominated by a toroidal field of a transverse profile,
  $b_\phi\propto R^{-1/2}$. The toroidal field dominated region covers
  the outer parts of the jet ($10R_{\rm L}\simless R\leq R_{\rm j}\simeq36R_{\rm L}\simeq3.6\times10^8$~cm) and the inner cocoon
  ($R_{\rm j}\simless R\simless R_{\rm ic}\simeq60R_{\rm L}\simeq
  6\times10^8$~cm). The outer cocoon, which contains the shocked
  ambient matter and is dominated by thermal pressure, covers the
  region $R_{\rm ic}\simless R\simless R_{\rm c}\simeq120R_{\rm L}\simeq
  1.2\times10^9$~cm.} 
\label{fig:3D_b_profiles} 
\end{figure}

To demonstrate the effect of the thermal pressure on the configuration of toroidal field, we return to the force balance equation (eq. \ref{eq:dbdr}) and include a thermal pressure term. In a cylindrical jet, the equation takes the form,
\begin{equation}\label{eq:dbdr_pg}
  \frac{db^2}{8\pi dR}+\frac{b_\phi^2}{4\pi R}+\frac{dp_{\rm th}}{dR}=0,
\end{equation}
where we neglect here the poloidal hoop stress term.  We assume an 
approximate equipartition between the thermal and magnetic pressures,
i.e. 
$p_{\rm th}=b^2/8\pi$.
  The solution to eq. \eqref{eq:dbdr_pg} inside the core where $b_\phi\ll b_p$, is $b_p={\rm const}$. The resultant profile of $b_\phi$ is
\begin{equation}\label{eq:b_phi_3D}
  b_\phi\propto
\left\{
\begin{array}{ll}
 R & \mbox{,  $R<R_0$}, \\
 R^{-1/2} & \mbox{, $R>R_0$}.  \ 
\end{array}
\right.
\end{equation}
This profile is similar to that of the headless jet
(eqs.~\ref{eq:b_phi_inf} and~\ref{eq:b_p_inf}); the only difference
here is that outside the core, $R > R_0$,
the high thermal pressure replaces $b_p$ and flattens out the
transverse profile of the toroidal field.
Figure \ref{fig:3D_b_profiles} shows the magnetic and
thermal pressure profiles across the simulated jet, above the dissipation
region. The profiles are measured in a snapshot of a jet in the model
M3, at a time $t=4400R_{\rm L}/c\simeq1.5$~s (same as in
Fig.~\ref{fig:3D_jet_fieldline}) and an altitude $z=400R_{\rm L}\simeq4\times10^9$~cm, which is well above the region where most of the dissipation takes place. We average the profiles in the azimuthal direction 
to smooth out the small axial asymmetries due to the external kink modes.
We show the azimuthally averaged $\avg{b_\phi}$ (solid  blue), $\avg{b_p}$ (dashed green), $\avg{b}$ (solid  grey) and $\avg{\sqrt{8\pi p_{th}}}$ (dash-dotted red).  As expected from eq. \eqref{eq:b_phi_3D}, the magnetic field configuration is split into two regions: i) an inner core, dominated by poloidal field with a flat profile that extends to $R_0\simeq10$, ii) an outer region, dominated by
toroidal field with a profile $b_\phi\propto R^{-1/2}$ that covers the
outer jet, at $R_0\lesssim R\leq R_{\rm j}\simeq 36R_{\rm L}\simeq3.6\times10^8$~cm, and extending into the inner cocoon, $R_{\rm j}<R\leq R_{\rm ic}\simeq 60\simeq6\times10^8$~cm. Within the jet, magnetic and thermal energies are indeed in equipartition, which is the assumption under which we obtained the scaling~\eqref{eq:b_phi_3D}. 

\begin{figure}
   \begin{center}
  \includegraphics[width=\columnwidth]{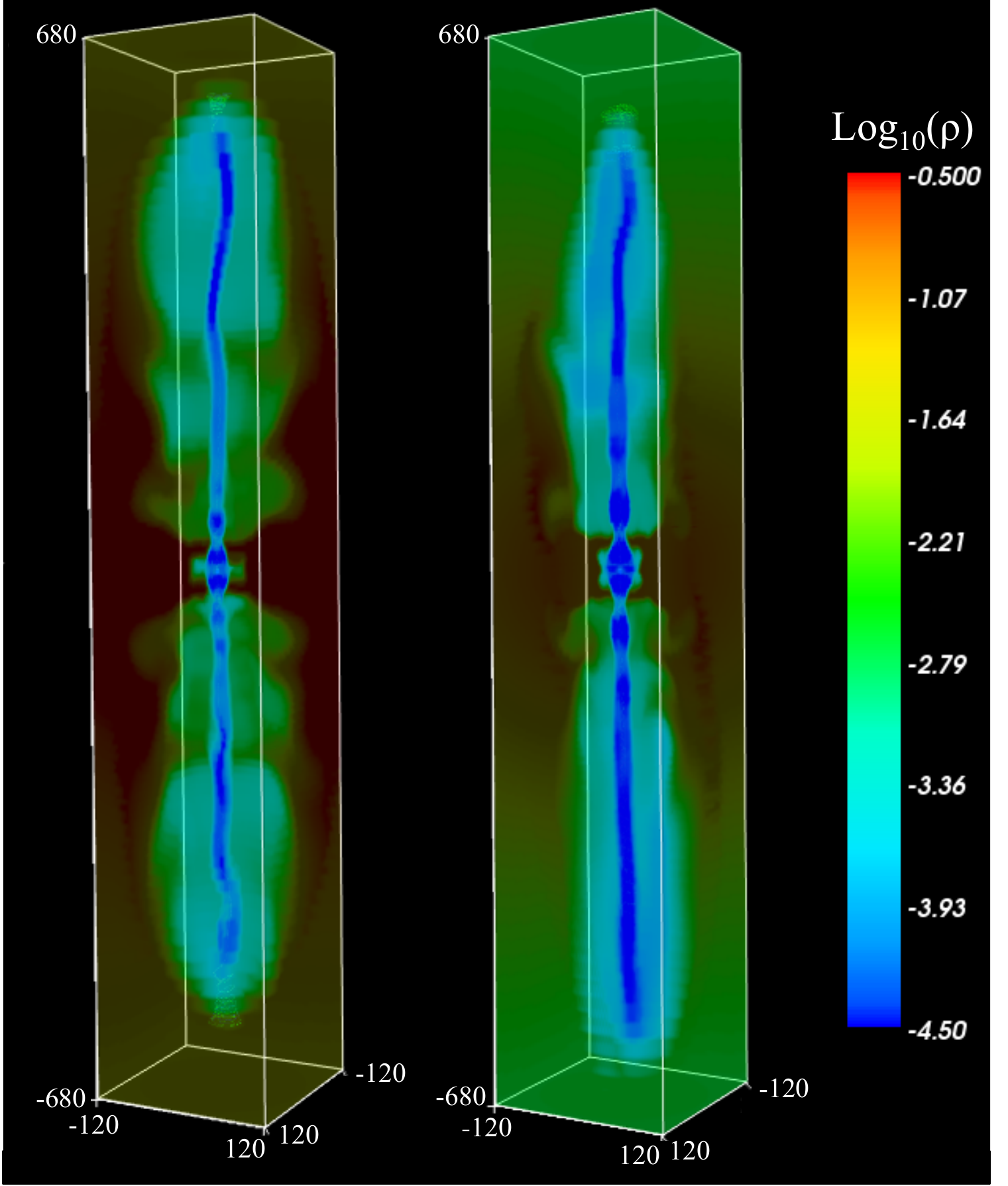}  
 \end{center}
 \caption{The morphology of the jets of different powers.  The left
   panel shows a low-power jet in model M3LP at
   $t=10995R_{\rm L}/c\sim3.7$~s. The right panel shows a high-power jet in
   model M3 at $t=4084R_{\rm L}/c\sim1.36$~s. The high-power jet is 10 times
   more power than the lower-power jet and propagates approximately
   $2.7$ times faster, and it reaches the same distance as the low-power jet
   in a shorter time. The large-scale bends of the jets are
   charactersistic of the external kink instability. 
   The low-power jet is thinner and becomes
   external-kink unstable at a smaller distance. As a result, kink
   modes grow to a larger amplitude and deform it more strongly than
   the high-power jet.
}
\label{fig:3dextkink}
\end{figure}

The combination of a flat magnetic field profile and a high thermal
pressure stabilises the jet. \citet{2012ApJ...757...16M} found that
kink instability is suppressed in jets with toroidal field profiles as
shallow as $\propto R^{-1/2}$, while \citet{mignone_3d_jets_2010}
reached a similar conclusion when they tested the stability of hot magnetised jets. Thus, above the dissipation region
the hot jet material is stable to the internal kink modes and is
expected to behave in a similar way to a hot, purely hydrodynamic
flow.
However, we find that kink instability can still grow around
the jet perimeter. This is known as the \emph{external kink} mode, which
produces helical motions that deform the entire body of the jet. This
is seen in Fig.~\ref{fig:3dextkink} as the large-scale bends of the
(blue) jets.  External kink grows on a time scale of the order of the
time it takes an Alfv\'{e}n wave to travel around the jet
perimeter. Typically it takes $\sim5-10$ growth times for the global
kink modes to grow substantially, and this is why most of the deformation occurs at
the top, ``older'' parts of the jet, mainly near the jet head
where the kink instability has the longest time to evolve. The helical
motions of the kink-unstable head increase the effective cross-section
of the jet and reduce its propagation velocity. If the
external kink grows to a high enough amplitude, it can even disrupt the jet
and cause it to stall. 

\subsection{Dissipation rates and convergence tests}\label{sec:discussion}

One of the important results 
of this work is the efficient dissipation of
the magnetic field at and above the recollimation nozzle. This reduces
the strength of the toroidal magnetic field component and leads to
equipartition between thermal and magnetic energies. 
A similar transition in the magnetic field configuration was seen in
the simulations of twisted magnetic flux tubes that underwhent
dissipation via the internal kink
\citep[e.g.][]{2009A&A...506..913H,2011ApJ...729..101G,2015A&A...576A..37P}.  Once started, the dissipation in our simulation occurs over a length
scale of $\simless 10 R_{\rm j}$, which is consistent with the expected
growth time of the kink instability across the entire jet (see
e.g.~Fig.~\ref{fig:sigma_high}). This rate is also consistent with the
dissipation rate found in other simulations
\citep[e.g.][]{2012ApJ...757...16M, 2014arXiv1408.3318P}.

\begin{figure}
   \begin{center}
  \includegraphics[width=\columnwidth]{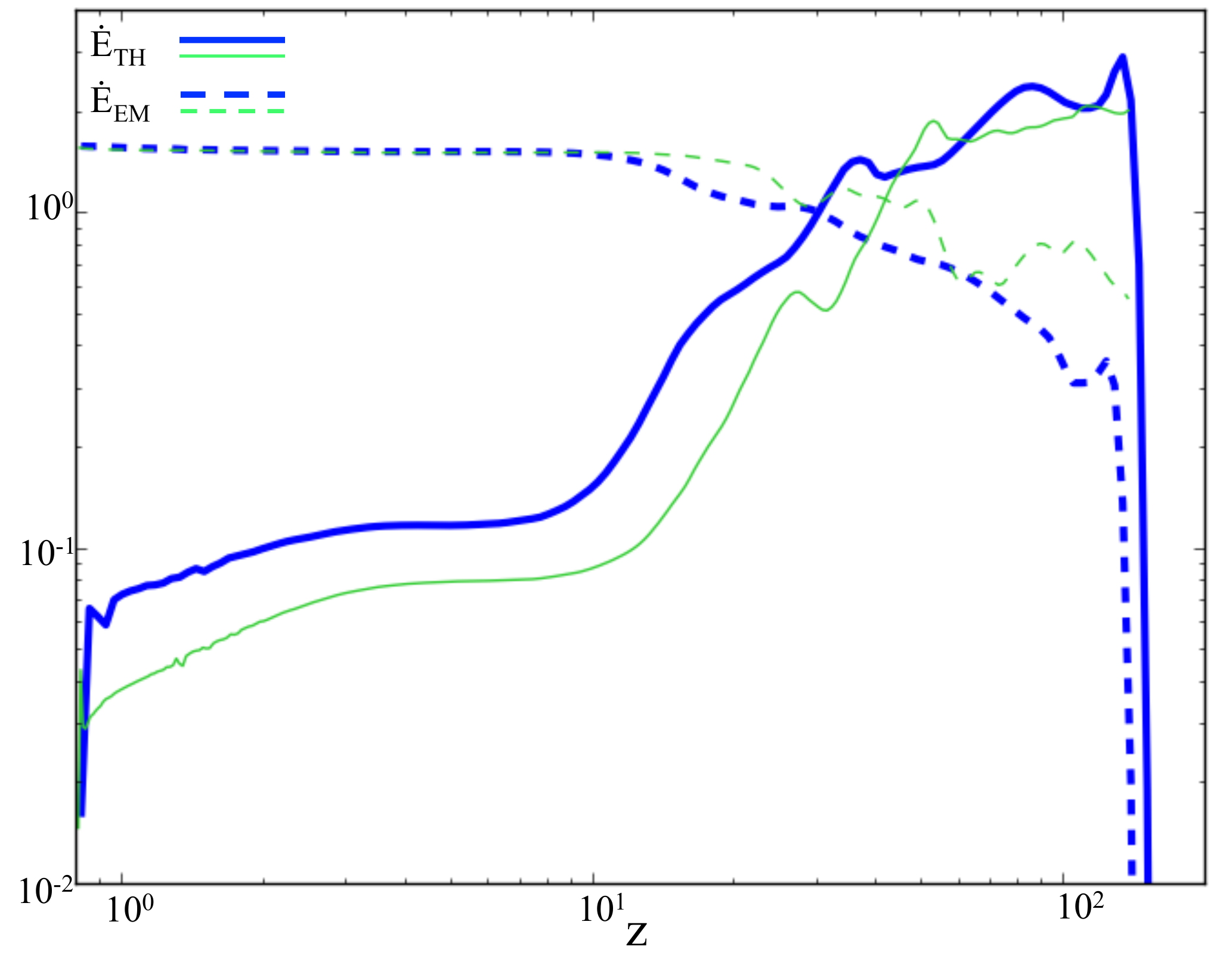} 
 \end{center}
 \caption{The thermal energy flux (solid lines) and the Poynting flux (dashed lines),  as a function of $z$. We show the fiducial model, M3, at $t=1885R_{\rm L}/c\sim0.6$~s (thick blue lines) and a
   high resolution model, M3HR, at $t=2515R_{\rm L}/c\sim0.8$~s (thin green lines). Both jets
   extend to the same distance, $z_{\rm h}\simeq150$, and have the same
   width, though the high resolution jet propagates at a velocity
   slower by $25\%$ due to somewhat increased wobbling motions of the
   head. Even though the jets show some differences at their heads, the
   dissipation rates at their recollimation points are essentially
   the same. This
   indicates that the dissipation could be driven by a large-scale
   instability and not affected by the microphysics.}
\label{fig:dissipation}
\end{figure}

To test the effect of the grid resolution on the dissipation rate, we
reran our fiducial simulation, M3, with twice the resolution in each
direction. We refer to this high-resolution simulation as model M3HR
(see Table~\ref{tab:models_grid}).  Figure \ref{fig:dissipation} shows
the electromagnetic energy flux (dashed lines) and the thermal energy
flux (solid lines) in the fiducial model M3 (blue)
and its high-resolution version, model M3HR (green).
For a proper comparison, the jets are taken to have the same
$z_{\rm h}$. Because in the high resolution simulation the jets propagate at
a velocity that is somewhat slower (by $\sim25\%$) than in our fiducial
resolution simulation, the snapshots are taken at slightly different
times. Even though the propagation velocities are
different, the dissipation rates are similar. Moreover, the jet width
in both cases is $R_{\rm j}\simeq 10R_{\rm L}\simeq10^8$~cm and the dissipation occurs over a
range $\Delta z\simless100R_{\rm L}\simeq10^9$~cm that agrees with the expected kink growth
timescale at this distance.
We have not tested how the dissipation rate depends
on the Riemann solvers used in the code \citep[see
e.g.][]{2012MNRAS.422.1436O}; this will be studied elsewhere. 

The $\sim25\%$ difference in the head velocity comes from the fact
that the high resolution jet is somewhat less stable to the external kink
than the lower resolution jet, and the wobbling motions of the head
have a somewhat larger radius. As we show in Sec. \ref{sec:analytical}, the head velocity is inverse proportional to the head radius. Thus the wobbling motions increase the effective jet radius by $\sim25\%$.
Our simulations are therefore not fully converged in the context of the external kink and the head velocity, and it is 
desirable to make a proper convergence test. Such a test, however, is hard to carry out in global simulations like ours, due to the large computational resources required. An alternative is to perform a convergence test for the
amplitude of the external kink instability in a small periodic box.
\citet{2009ApJ...700..684M} conducted such a test and concluded that
when the jet is resolved by more than 20 cells across, the amplitude of the external kink is not sensitive to the resolution.  In our case
the fiducial resolution jet contains $\sim14$ cells in the transverse
direction at $z_{\rm h}=160R_{\rm L}\simeq1.6\times10^9$~cm, where we make the comparison between the jets.
In the high resolution jet the number of cells is twice as large.
Thus it is likely that we are close to the full convergence.

To sum up, our results are rather insensitive to the numerical
resolution, suggesting the possibility that the dissipation rate and the jet
velocity are  determined by the underlying instability and not the
details of the numerical scheme. Whereas the quantitative
understanding of the dissipation rate requires even higher resolution
studies to probe the asymptotic scaling of the instability properties
with resolution, the saturation of the instability at equipartition
between thermal and magnetic energies is a robust, important result of this work.

\section{An analytic model for a hot magnetic jet}\label{sec:analytical}
In the previous sections, we saw that the jet material dissipates
about half of its magnetic energy via the internal kink instability. The dissipation continues until equipartition is reached between thermal and magnetic energies. Since the jet is hot and has a relatively
flat transverse pressure profile, it is stable to the internal kink
modes (see Fig.~\ref{fig:3D_b_profiles} and
Sec.~\ref{ssec:equipart}). However, as Fig.~\ref{fig:3dextkink} shows,
that the jet (seen in dark blue colour) develops large-scale bends,
characteristic of external kink modes. This means that, unlike hot
hydrodynamic jets, hot magnetised jet can be unstable to the external
kink mode. In the extreme cases, it is conceivable that the external kink
could even lead to a complete disruption of the jet. In our
simulations, we find that the external kink causes the head of the jet
to wobble sideways, increasing the  effective cross-section of the jet
and decreasing its propagation speed.

Here, we investigate the conditions under which 
external kink modes can grow in the jet to a level where they affect jet
propagation.  Motivated by the fact that the jet is hot (see
Sec.~\ref{ssec:equipart}), we make use of its similarity to a
hydrodynamic jet and employ an analytic model for the propagation of a
hydrodynamic jet in a medium by \citet{2011ApJ...740..100B}.  We
calculate the conditions in the jet and check if the external kink has
sufficient time to grow and considerably deform the jet. A reader
interested in the analytic results may skip directly to
eq.~\eqref{eq:Lambda}, which gives the growth rate, and
eq.~\eqref{eq:bh}, which estimates the propagation velocity of the
jet.

\begin{figure}
  \begin{center}
\includegraphics[width=\columnwidth]{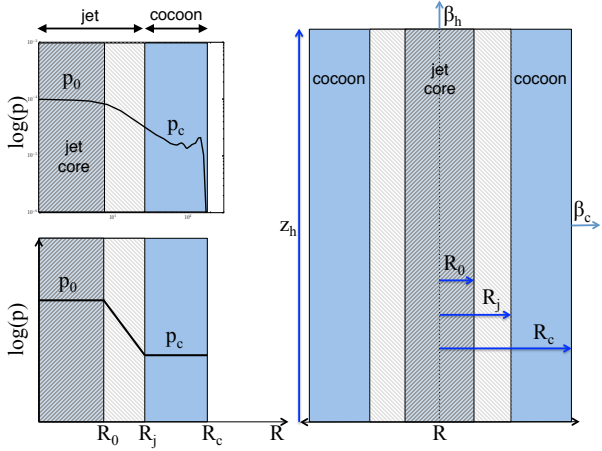}  
\caption{A schematic representation of our analytic jet model. Right panel: the
  jet (shown in grey) is composed of an inner poloidal--field--dominated
  core of radius $R_0$, surrounded by a toroidal--field--dominated
  sheath of radius $R_{\rm j}$. The jet is
  engulfed by a cocoon (light blue) of radius $R_{\rm c}$. The jet head, located at $z=z_{\rm h}$,
  propagates at the velocity $\beta_{\rm h}c$ and the cocoon expands
  laterally at the velocity $\beta_{\rm c} c$.  Upper left panel: A lateral
  total pressure profile in a slice through the midpoint of
  a 2D jet in model M2 at time $t=730R_{\rm L}/c\simeq2.4$~s.  Lower left panel:
  an analytic approximation for the pressure profile. We assume a flat
  pressure profile in the jet core and the cocoon, and in the sheath
  we approximate the pressure as $p\propto R^{-1}$.}
\label{fig:jet_cocoon_model}
\end{center}
\end{figure}
 
Figure \ref{fig:jet_cocoon_model} illustrates our analytic jet
model.  We approximate the jet and the cocoon as concentric cylinders
of the same height, $\zh$, and of radii $R_{\rm j}$ and $R_{\rm c}$, respectively.  We
assume that the ambient medium is cold and has a power law 
density profile, $\rho_a\propto z^{-\alpha}$.  Here and below, we use the
indices $j, h, c$ and $a$ to describe quantities in the jet, the jet head, the cocoon and the ambient medium, respectively.

The jet and the cocoon are in pressure balance at their interface, $R=R_{\rm j}$. We approximate the pressure in the cocoon as uniform, thus it can be evaluated as  
\begin{equation}
   p_c\simeq \frac{E_c}{3V_c}
\label{eq:pc0}
\end{equation}
where $E_c$ is the total energy in the cocoon, $V_c$ is the cocoon
volume, and we assume here a relativistic thermal equation of
state. 

The cocoon energy is injected at the jet head (see Sec.~\ref{sec:2D}). Its value can be estimated as
\begin{equation}
E_c=\int L_{\rm j}(\beta_{\rm j}-\beta_{\rm h})dt,
\label{eq:Ec}
\end{equation}
where $L_{\rm j}$ is the jet luminosity, $\beta_{\rm j}$ and $\beta_{\rm h}$ are the velocities of the jet material and the head, respectively. The factor $(\beta_{\rm j}-\beta_{\rm h})$ measures the
reduction in the energy flux due to the motion of the head. The volume of
the cocoon is estimated as
\begin{equation}
V_c\simeq\pi c^3\int\beta_{\rm h}dt\left(\int\beta_{\rm c}dt\right)^2,
\label{eq:Vc}
\end{equation}
where $\beta_{\rm c}=\sqrt{p_c/\rho_ac^2}$ is the sideways expansion
velocity of the cocoon due to its own pressure. 
 Combining
eqs.~\eqref{eq:pc0}, \eqref{eq:Ec}, and \eqref{eq:Vc}, and taking $t=\int dz_{\rm h}/\beta_{\rm h}c$ we obtain (up to
an order unity numerical factor):
 \begin{equation}\label{eq:pc}
   p_c\simeq\fracb{L_{\rm j}^3\rho_a}{\pi^3z_{\rm h}^4R_{\rm j}^4c}^{1/4},
 \end{equation}
where $\rho_a$ is evaluated at the jet head. A full calculation of the pressure  is presented in Appendix~\ref{sec:analytic-model-jet}, where we follow the method of \citet{2011ApJ...740..100B}.

Figure \ref{fig:jet_cocoon_model} shows a schematic plot of the pressure profile in the jet. Using the scaling given in
Sec. \ref{ssec:equipart} and eq. \eqref{eq:b_phi_3D} we get:
\begin{equation}\label{eq:Pj}
   p_j=
\left\{
\begin{array}{ll}
 p_0 & \mbox{, $r\leq R_0$},\\
 p_0\displaystyle\left(\frac{R_0}{R}\right) & \mbox{, $r>R_0$},
\end{array}
\right.
 \end{equation}
 where we take $R_0\simeq R_{\rm j}/2$ as the radius of the poloidal field dominated core. The jet pressure is connected to its luminosity
 via 
 \begin{equation}
L_{\rm j}=2\pi\int_0^{R_{\rm j}}
\frac{\Gamma}{\Gamma-1}p_j\gamma_{\rm j}^2\beta_{\rm j}cRdR,
\label{eq:Lj}
\end{equation}
 where $\Gamma$ is the adiabatic index. An equipartition between the magnetic and thermal pressures results in  $\Gamma=3/2$ (See Appendix~\ref{sec:analytic-model-jet}).
   Substituting eq.~\eqref{eq:Pj} into eq.~\eqref{eq:Lj}, we get the jet pressure at $R_{\rm j}$: \begin{equation}\label{eq:pj_Rj}
  p_j(R_{\rm j})\simeq\frac{2L_{\rm j}}{9\pi\gamma_{\rm j}^2\beta_{\rm j}R_{\rm j}^2c}
\end{equation} 
We can now use eqs.~\eqref{eq:pc} and \eqref{eq:pj_Rj} to solve for $\gamma_{\rm j}R_{\rm j}$ from the condition of pressure balance at the at the edge of the jet:
 \begin{equation}\label{eq:rjgj}
   R_{\rm j}\gamma_{\rm j}\simeq\sqrt{\frac{2}{9\pi\beta_{\rm j}}}\fracb{L_{\rm j}\zh^4}{\rho c^3
   \gamma_{\rm j}^2}^{1/6}\Psi
 \end{equation}
 where 
 $\Psi=\left[\pi(5-\alpha)(3-\alpha)/6\right]^{1/3}$ is an integration constant of order unity (see Appendix~\ref{sec:analytic-model-jet}).

External kink modes grow in the jet on a time scale of the order of the
Alfv\'{e}n crossing time around the jet perimeter (eq.~\ref{eq:t_kink}):
\begin{equation}\label{eq:t_kink_ext}
   t_{\rm kink}\simeq \frac{2\pi R_{\rm j}\gamma_{\rm j}}{v_{\rm A}},
\end{equation}
where we made use of the fact that toroidal and poloidal fields are
in equipartition at the jet edge.
As we discussed above, typically it takes $\sim 5{-}10\times t_{\rm kink}$ for the
instability to develop to a level where the jet is considerably
deformed \citep{2012ApJ...757...16M}. The time
available for the instability to grow is the time it takes a fluid
element to reach the jet head:
\begin{equation}
t_{\rm dyn}\simeq \frac{z_{\rm h}}{c(\beta_{\rm j}-\beta_{\rm h})}\simeq\frac{z_{\rm h}}{c\beta_{\rm j}}, 
\label{eq:dyn}
\end{equation}
where we work in the limit where $\beta_{\rm h}\ll\beta_{\rm j}$.
The jet
will therefore be considerably kinked if
\begin{equation}\label{eq:tk_t_dyn}
  \frac{10\eta t_{\rm kink}}{t_{\rm dyn}}\simeq\frac{20\pi  R_{\rm j}\gamma_{\rm j}\beta_{\rm j}}{z_{\rm h}}\lesssim1,
\end{equation}
where $0.5\simless \eta\simless1$, and we made
use of the fact that the jet is highly magnetised, i.e., that
$v_{\rm A}\approx c$. Substituting $R_{\rm j}\gamma_{\rm j}$ from
eq. \eqref{eq:rjgj} we get an expression for the stability of the jet
to the external kink mode:
  \begin{equation}\label{eq:Lambda}
   \Lambda\equiv\frac{10\eta t_{\rm kink}}{t_h}\simeq20\sqrt{\frac{2\pi}{9}}\fracb{L_{\rm j}}{\rho_a\zh^2\gamma_{\rm j}^2c^3}^{1/6}
   \frac{\sqrt{\beta_{\rm j}}}{\beta_A}\eta\Psi
 \end{equation}

\begin{figure}
   \begin{center}
  \includegraphics[width=\columnwidth]{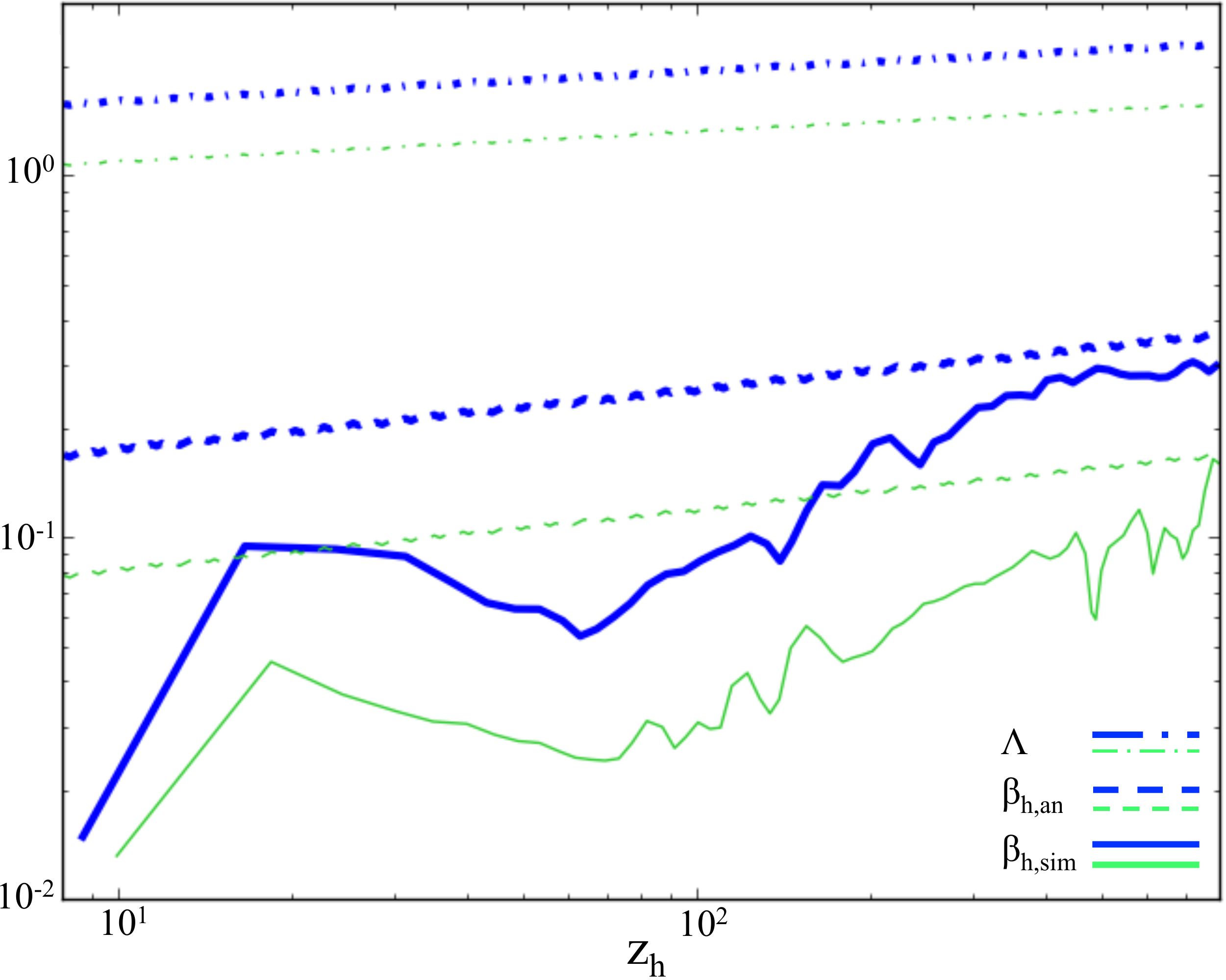} 
 \end{center}
 \caption{Jet head velocity $\beta_{\rm h}$ as a function of head position
   $z_{\rm h}$ for high-power (thick blue lines, model M3) and low-power (thin green lines, model
   M3LP) jets. The high-power jet is a factor $10$ more luminous and
   propagates a factor of $\sim 2$ faster than
   the low-power one. We show the simulated values of $\beta_{\rm h}$ with
   solid lines, the analytic approximations according to
   eq.~\eqref{eq:bh} with dashed lines, and the corresponding $\Lambda$
   parameters from eq.~\eqref{eq:Lambda} with dash-dotted lines. Once
   $\Lambda\gtrsim2$, the velocity profile approaches the analytic approximation, indicating the the head wobbling motions are small, and the jet propagates similarly to a hydrodynamic jet.}
\label{fig:sim_vs_model}
\end{figure}
The propagation velocity of the jet is directly linked with the cross
section of the jet head. In the limit where the head velocity is
non-relativistic ($\gamma_h\beta_{\rm h}\simless1$), we can write \citep{2003MNRAS.345..575M,2011ApJ...740..100B}:
 \begin{equation}
  \beta_{\rm h}\simeq\sqrt{\frac{L_{\rm j}}{\pi R_h^2\rho_a c^3}},
\label{eq:betah}
\end{equation}
where $R_h$ is the cross sectional radius of the jet head. As long as the 
helical motions of the head, due to the kink instability, are not large enough to increase the cross section of the jet substantially, we can use eq. \eqref{eq:rjgj} to estimate the velocity of the head:
\begin{equation}\label{eq:bh}
  \beta_{h}\simeq\sqrt{4.5\beta_{\rm j}}\fracb{L_{\rm j}\gamma_{\rm j}^4}{\zh^2\rho_ac^3}^{1/3}\Psi^{-1}
\end{equation}  
Figure \ref{fig:sim_vs_model} shows $\beta_{\rm h}$ as a function of the
location of the jet head in models M3 and M3LP,
representing a GRB jet with a typical luminosity and with a low luminosity, 
respectively. We compare the values of $\beta_{\rm h}$ measured from the
simulations, shown with thick solid lines, to the analytic values
due to eq.~\eqref{eq:bh}, shown with thin dashed lines.
Values for the high (low) luminosity jet are shown in blue (red). We
also show the values of $\Lambda$ from eq. \eqref{eq:Lambda} in
dash-dotted lines. Both jets are marginally stable at small $z_{\rm h}$ (early
times), and become increasingly more stable with time. 

In the high-luminosity jet, the wobbling motions of the head are relatively large
at $z_{\rm h}\simless300$. In this regime, $\beta_{\rm h}$ deviates from the
analytic solution and has a steeper dependence on $z_{\rm h}$. As the jet
head propagates to higher altitudes it becomes more stable and the wobbling amplitude
decreases. The effective cross-section of the jet decreases with
increasing $z_{\rm h}$, thus the head velocity increases at a faster rate than the analytic expectation.
At $z\simgreat 300$ the simulated velocity profile is close to the analytic
approximation (to within $\sim10\%$) and exhibits a similar profile,
$\beta_{\rm h}\propto z_{\rm h}^{1/6}$, reflecting the increased stability of the
head to kink modes. 
In the low-luminosity jet, on the other
hand, the wobbling motions remain large until the head reaches the altitude
$\abs{z}=800R_{\rm L}\simeq 8\times10^9$~cm. Therefore the velocity remains below the analytic
approximation and follows a steeper $z_{\rm h}-$dependence. Based on these two cases, we suggest that the simulated $\beta_{\rm h}$ approaches the analytic solution at $\Lambda\simgreat2$.

Note that for both high- and low-luminosity jets, at
$z_{\rm h}\simless 60$ the head velocity decreases with increasing $z_{\rm h}$. This
decrease represents early times in the simulation when the jet head
just starts to propagate outward and the magnetic field on
the surface of the NS is very high due to the buildup of the toroidal
magnetic field
when the jet was just formed. This leads to an increase in the Pointing
flux which pushes the head at a greater force. As the jet begins to
propagate, the excess of
$B_\phi$ and the Poynting flux decrease. 

Equations \eqref{eq:Lambda} and \eqref{eq:bh} depend on the
inverse of $(\rho_a z_{\rm h}^2)$. Since this expression decreases in
density profiles steeper than $z^{-2}$, in such profiles the jet
accelerates and becomes progressively more stable to kink modes with
increasing distance.
In contrast, in density profiles shallower than $z^{-2}$, the
jet would decelerate and become less stable with the increasing
distance from the central compact object. In this case,
the propagation velocity would be slower than the one obtained in
eq.~\eqref{eq:bh}, since the wobbling of the jet head would increase the
effective cross-section of the jet. Therefore, the jet would have to push a larger
amount of gas to propagate. In such a case, we expect that there would be some altitude at which the head velocity will be comparable to the
expansion velocity of the cocoon, $\sqrt{p_c/\rho_a}$. When this
happens, the cocoon would begin to overtake the jet, and we expect
that the jet
would stall.

\section {Astrophysical Applications}\label{sec:astrophysical}
  
We now consider our analytic model for the jet propagation
in the context of two systems that are
expected to host magnetised relativistic jets: GRBs and AGN. In the
GRB case, we take a jet with a typical power of $3\times10^{49}$ erg/s
\citep[e.g.][]{2005ApJ...619..412G}, that propagates in a fiducial
Wolf-Rayet star of mass $M_*=10M_\odot$, radius
$r_{_*}=10^{11}$ cm, and density profile
$\rho_a\propto r^{-\alpha}$, with $\alpha=2.5$. The stability
parameter (eq. \ref{eq:Lambda}) of the jet is
\begin{equation}\label{eq:Lambda_GRB}
   \Lambda_{\rm GRB}(\zh)\simeq2.6\fracb{\zh}{r_*}^{(\alpha-2)/6}\fracb{L_{{49.5}}r_{{*,11}}}{M_{{*,34.3}}\gamma_{{0.3}}^2}^{1/6}\fracb{(5-\alpha)^2(3-\alpha)}{3.125}^{1/6}\eta, 
\end{equation}
where $\eta\sim0.5{-}1$ (see eq.~\ref{eq:tk_t_dyn}) and
 we take $\sqrt{\beta_{\rm j}}/\beta_A\sim1$. Here, we used the
notation $A_x\equiv A\times10^{x}$. The inefficient acceleration of
the jet material below the head implies that $\gamma_{\rm j}$ changes only
weakly as a function of the position of the jet head and remains
approximately $2$ over many decades. Eq.~\eqref{eq:Lambda_GRB},
therefore, gives $\Lambda_{\rm GRB}\approx2.6$, which suggests that a GRB
jet is expected to be only marginally deformed by the external kink
instability as it propagates through the star, and that the jet head
velocity should not deviate much from the analytic approximation~\eqref{eq:betah}
for $\beta_{\rm h}$. The weak dependence of $\Lambda$ on $\zh$ implies that even
close to the light cylinder, at $\zh=10^{7}$ cm, $\Lambda_{\rm GRB}$ remains larger than $1$, thus the
jet is expected to be marginally stable also in the stellar interior.
However, as we see from the simulations, 
the marginal stability implies substantial wobbling motions 
that reduce its
propagation velocity to below the analytic
expectation~\eqref{eq:betah}, in regions with $\Lambda\simeq1$. 
By substituting the GRB parameters into eq. \eqref{eq:bh}, we obtain
an estimate of the jet head velocity,
\begin{equation}\label{eq:bh_GRB}
  \beta_{\rm h,GRB}(\zh)\simeq 0.3\fracb{\zh}{r_*}^{(\alpha-2)/3}\fracb{L_{{49.5}}r_{{*,11}}\gamma_{{0.3}}^4}{M_{{*,34.3}}}^{{1}/{3}}\fracb{(5-\alpha)(3-\alpha)^2}{0.625}^{-{1}/{3}},
\end{equation}
which as we show in Sec. \ref{sec:analytical}, provides a good approximation for the true velocity outside $r\simeq3\times10^9$~cm, or outside of the inner few percents of the stellar radius where $\Lambda\simgreat2$. 
Thus, the propagation velocity of a typical GRB jet in the star is
sub-relativistic, and well described by eq.~\eqref{eq:bh_GRB}.

The breakout time of the jet from the star can be obtained by
integrating $\int_{_0}^{{R_*}} dz/\beta_{\rm h} c$, which gives
$t_b\simeq 10$ sec for the fiducial parameters used here.
\citet{2012ApJ...749..110B,2014MNRAS.443.1532B} showed that the
distribution of GRB duration times fits a typical breakout time of the
jet from the progenitor star of about $10$ sec, in agreement with the
breakout time we obtain here. Note that this propagation velocity is
somewhat slower than the propagation velocity of an equivalent 
hydrodynamic jet obtained analytically
\citep{2011ApJ...740..100B} and in 2D
simulations \citep[e.g.][]{mor06,2009ApJ...699.1261M}. Although recent
3D simulations of hydrodynamic jets indicate that the two velocities
are actually similar \citep{2013ApJ...767...19L}.
        
When $\Lambda<1$ the wobbling motions of the jet head due to the
external kink are large enough to decrease $\beta_{\rm h}$ significantly. As
a result, the breakout time of the jet from the star becomes much
longer than $10$ sec. A typical core-collapse GRB engine is active for
at least as long as the GRB lasts \citep{1997ApJ...485..270S}, or a
few tens of seconds \citep{kouv93}. If the jet power decreases, the
jet becomes less stable and slower, and chances are that the central
engine will shut off before the jet head reaches the stellar edge. In
this case, the jet will not be able to break out of the star and will
remain buried in it. 

Such failed jets are favourable sources for
producing low-luminosity GRBs that could occur when the cocoon breaks
out of the stellar surface
\citep{1998Natur.395..663K, 2001ApJ...550..410M, 2001ApJ...551..946T,
  2007ApJ...664.1026W, 2007ApJ...667..351W, 2010ApJ...716..781K,
  2011ApJ...739L..55B, 2012ApJ...747...88N,2015ApJ...807..172N}. The minimum power below
which the jet becomes kink-unstable in the star is obtained by setting
$\Lambda_{\rm GRB}(r_*)=1$ in eq.~\eqref{eq:Lambda_GRB}, giving a (single) jet power of
\begin{equation}\label{eq:Lc_GRB}
   L_{\rm min}^{{\rm GRB}} \simeq 10^{47}M_{{*,34.3}}r_{{*,11}}^{-1} \gamma_{{0.3}}^2\fracb{3.125}{(5-\alpha)^2(3-\alpha)}\eta^6~ \rm{erg/sec}.
 \end{equation}
The associated isotropic equivalent luminosity is $L_{\rm min,iso}^{{\rm GRB}} \simeq 2.6\times10^{49} (\theta/7^{\circ})^{-2}$ {erg/sec}, assuming a jet with a characteristic half-opening angle of $10^{\circ}$.
A jet with a lower power is
less likely to produce a regular GRB (see also \citealt{2007RSPTA.365.1129W}). Interestingly, the observed
distribution of GRB luminosities shows a cutoff at isotropic
equivalent luminosities lower than $\sim3\times10^{49}$ erg/sec
\citep[e.g.][]{2011MNRAS.416.2174C}. A slow, kink-unstable jet,
provides a natural explanation for such a cutoff.

After the jet breaks out from the star, the jet head accelerates and
loses causal contact with the jet base. The jet, in principle,
should relax back to the configuration of a steady state, headless jet
that is relatively stable to kink modes and is less likely to undergo
internal energy dissipation, as we discussed in
Sec.~\ref{sec:ideal}. However, at this point the jet is sufficiently
wide so that the collimation point  is located
far outside the light cylinder. In such a case, the poloidal field is too
weak to counterbalance the contracting force of the hoop stress,
resulting in a large pinching of the jet at the collimation
nozzle (see Sec. \ref{ssec:internal_kink}). The converging field lines have a large dispersion of pitch angles, and they converge into a small enough region that internal kink modes can grow and efficiently dissipate the magnetic energy, even without the extra compression of toroidal field coming from the jet head. We have run 
a version of our fiducial model, M3, in which the jet breaks out of a ``stellar surface'' at $r=800R_{\rm L}$ and verified that the dissipation at the recollimation nozzle continues long after the breakout. A full analysis of the  jet at the post breakout stage will be done in the future.

The jet material that emerges from the star is therefore
relativistically hot, with half of its energy in thermal pressure and
half still locked in the form of magnetic field.  This means that
there is sufficient thermal and magnetic energy to accelerate the jet
material to high Lorentz factors that are inferred from the
observations \citep[see,
e.g.,][]{lithwick_lower_limits_2001,pk02}. The jet width at that time
is of the order $R_{\rm j}\sim10^9$ cm (eq. \ref{eq:rjgj}) and agrees to an
order of magnitude with the inferred size of the emission region in
GRB~970828 \citep{2007ApJ...664L...1P}. The site of the
dissipation, which takes place at about $z\sim10R_{\rm j}\sim10^{10}$~cm, is
located sufficiently deep in the star, so that the amount of thermal
photons required by observations of prompt emissions can be produced
\citep{2013ApJ...764..143V}. Note that the generation of the high
energy non-thermal emission, seen in many GRBs, potentially requires an additional
dissipation process to take place relatively close to the photosphere,
which can lie outside of the star
\citep[e.g.][]{1990ApJ...365L..55S,2000ApJ...530..292M,2011ApJ...733...85B}. This
topic is beyond the present work.

Our results have important implications for AGN jet classification which divides AGN radio galaxies into two types: Fanaroff-Riley type I and II (FRI and FRII) galaxies \citep{1974MNRAS.167P..31F}. FRII galaxies show stable jets that typically extend over several hundred kiloparsecs
\citep[e.g.][]{1998MNRAS.296..445H,2013ApJ...767...12G}, with a
well-defined hotspot at the jet head, and a cocoon that extends from
the jet head backwards. The jets in FRI galaxies, on the other hand,
are much shorter on average, extending over several tens of kiloparsec
\citep[e.g.][]{2000AJ....120.2950X}. They emit radiation throughout
the jet body, indicative of volumetric dissipation, and their cocoons
usually outrun the jet heads. A similar
morphology to that of the FRI galaxy jets is expected when the jet head
becomes kink unstable and its velocity drops below the expansion
velocity of the cocoon. As we show in Sec.~\ref{sec:analytical}, this
can occur if the jet propagates in a medium with a density profile
that is flatter than $r^{-2}$.  Recent studies have shown that the
density profiles in clusters of galaxies are relatively flat with
$\rho_a\propto r^{-1}$ at the inner part of the cluster
($r\simless 100$ kpc) and steepen to $\rho_a\propto r^{-2}$
further out \citep[e.g.][]{2013ApJ...765...24N}. In this case, a jet with a
luminosity below
\begin{equation}
  L_{\rm min}^{\rm FR}\simeq4\times10^{45} \frac{\rho_a}{10^{-26}{\tiny\frac{\rm g}{\rm cm^3}}}\fracb{r}{100~ \small{\rm kpc}}^2\fracb{\gamma_{\rm j}}{2}^2 \fracb{3.125}{(5-\alpha)^2(3-\alpha)}\eta^6~ {\rm erg\;s^{-1}}
\end{equation}
will have a stability parameter $\Lambda<1$ at $r<100$~kpc
and will become unstable to external kink before leaving the inner flat density core.  The wobbling motions increase the effective jet cross-section of the jet and reduce the propagation velocity until it becomes comparable to the expansion velocity of the cocoon. From this point on, the jet is effectively stalled. The continuous injection of energy at the jet head results in a prominent, massive cocoon relative to the jet. In addition, the increasing wobbling motions can trigger dissipation along the jets that leads to radiation, making the jets shine and appear as FRI jets. 
In the opposite case, when $L>L_{\rm min}^{\rm FR}$, the jets are able to break out
of the core before they become unstable. Once out, they accelerate in
the steep density profile outside the galaxy core and become increasingly
stable. These jets are expected to have more elongated, less energetic
cocoons, and are expected to have well-defined working surfaces at
their heads where most of the work is being done to push the ambient
medium. Such a morphology is observed in FRII jets.

\section{Comparison with other works}\label{sec:other_works}

The stability of magnetised jets was studied extensively via 3D
simulations, both in local, periodic boxes and in global simulations.
In many studies, the jet morphology and the magnetic field
configuration (including the pitch angle, or toroidal to poloidal field strength
ratio) is prescribed in an ad hoc way. In addition, the jets are
usually assumed to be of a cylindrical shape, and some initial perturbations
are imposed on the field lines to break the axial symmetry and
jump-start the kink instability. This suggests that in such studies
jet stability can be sensitive to the details of the initial jet setup
(e.g., the parameters of jet injection boundary conditions, the
initial conditions with which the jets are initialised, etc.) 
making it difficult to draw firm conclusions about jet stability.
By comparing the properties of our self-consistently launched jets to
other works, we can build up intuition on what effect the particular
simulation details have on the simulation outcome and where our
results fall in the global landscape of simulated jet configurations.
  
\citet{2012ApJ...757...16M} studied the stability of a stationary jet
with a poloidal field dominated core surrounded 
by toroidal field  dominated sheath, which extended to the edge of their grid. They tested
several configurations of toroidal and poloidal field lines in the
sheath which characterised by different values of $b_\phi/b_p$. They found that the instability grew faster for steep lateral field profiles and
higher ratios of $b_\phi/b_p$. When the lateral toroidal profile in the
sheath became flatter than $R^{-1/2}$, the instability grew over a longer timescale and eventually saturated. %

\citet{2014arXiv1408.3318P} conducted a similar study using jets that
contained only toroidal magnetic field with a profile $b_\phi\propto
R^{-1}$ outside a core. They replaced the poloidal field with thermal pressure, which
dominated the core, was sub-dominant outside, and had a flat profile
outside the core (they found that the inclusion of poloidal field had
no noticeable stabilising effect). Their
jets, which are moderately magnetised, became kink unstable over a time scale that is consistent with the expected growth time, $\sim10t_{\rm kink}$. They witnessed that the energy dissipation associated with the kink
instability led to an increase in the jet core size, similar to our
findings. Since their jets were infinite, the instability could grow in the jet indefinitely, and the jets were eventually 
disrupted after $\sim40t_{\rm kink}$.  They also confirmed that the jets
remain stable if they expand sideways with a velocity that is fast
enough, so that the conditions for strong transverse causal contact are not met.
We find that in a time-dependent case, the sideways expansion of the cocoon
does lead to an increase in the jet
radius with time. However, in the environments that are relevant for
GRB and AGN jets, the rate of the expnasion is slow, and the jet
remains in strong causal contact.

Studies of the energy dissipation processes in stationary, non-relativistic
magnetised jets were conducted by
\citet[][]{2009A&A...506..913H,2011ApJ...729..101G} and
\citet{2015A&A...576A..37P} in the context of solar flares. They all
considered a cylindrical jet with poloidal field dominated core
surrounded by a toroidal field dominated sheath with a profile steeper
than $b_\phi \propto R^{-1}$. In their setup the jet is surrounded by a uniform magnetic field in
the $z-$direction, which provides support for the jet from the sides,
similar to the effect of the cocoon in our case. In all of their
simulations the internal kink evolved in the jet on a time scale 
comparable to the expected linear growth growth time, $\sim10t_{\rm kink}$.
The internal kink resulted in magnetic energy dissipation over a comparable time scale.
It led to the dissipation of most of the toroidal
component of the field, and to a jet with mostly a poloidal field. This
result is consistent with the internal dissipation in our collimated jets.

Global simulations of Poynting dominated jets were
conducted by \citet{2013MNRAS.436.1102M}. They injected a cylindrical
jet at the lower boundary of their grid and tracked its propagation.
The ambient medium was composed of a spherically symmetric supernova gas that expands homologously into a uniform density medium. 
The jets were assumed to be cylindrical, with only toroidal field
and thermal pressure. The thermal pressure dominated in the core
($R_0\sim1/3 R_{\rm j}$), and the toroidal pressure dominated outside the
core. They tested jets with several $\sigma$ parameters and Lorentz
factors. Their jets were deformed considerably above
$\sim10t_{\rm kink}c$, as expected. They also found indications that at the
region where magnetic energy was dissipated, field lines became more
aligned with the average jet velocity, indicating  an effective decrease of
toroidal field in the jet.  The overall morphology of the jet and its
propagation velocity were consistent with our results.

\citet{2007ApJ...656..721N} carried out global, non-relativistic MHD
simulations of AGN jets propagating in a gravitationally stratified
isothermal medium. They injected non-rotating, magnetised jets with a
prescribed magnetic pitch $b_\phi/b_p\sim25$. They found evidence of both
internal kink at smaller distances and external kink at larger
distances once they perturbed the ambient medium. They did not find
any jet heating by the internal kink. 

\citet{2014ApJ...781...48G} carried out relativistic MHD simulations
of non-rotating jets injected at a distance of $\sim10^3$
gravitational radii into a uniform background medium and followed them
for about $3$ orders of magnitude in distance. The transverse radius
of their injected jets was marginally resolved by $2{-}3$ grid cells.
They found that when the jet material is injected with 
 ratios of $b_\phi/b_p>10$ it develops internal kink modes and experiences substantial dissipation of magnetic energy, similar to our work. 
Above the dissipation regions, their jets seem to be considerably
more bodily kinked then ours. This is possibly because they propagate
in a uniform density profile that applies a larger resistive force on
the jet head than in our case. Similar to our work, they found that
the instabilities slowed down the jet propagation but did not disrupt
the jets.

On the analytic front, \citet{lyub09,2011PhRvE..83a6302L} obtained
steady state solutions to the structure of collimated Poynting
dominated jets. He showed that when the jet is collimated far from the
linear acceleration regime, it is focused into a recollimation nozzle
due to the excess force of hoop stress. Based on these solutions,
\citet{2010MNRAS.402..353L,2012MNRAS.427.1497L} postulated that if the recollimation
nozzle is narrow enough, strong dissipation should occur which will
effectively convert the magnetic into thermal energy. Models for the
propagation of a Poynting dominated GRB jet in a star were developed
by \citet{2013ApJ...764..148L} and \citet{2014MNRAS.443.1532B}. Both
groups obtained similar, mildly relativistic values for the velocity
of the jet head. However, since they ignored the wobbling motions of
the head due to the external kink instability, they overestimated the
head velocity. Based on their findings of the jet width,
\citet{2013ApJ...764..148L} concluded that the magnetic field is
dissipated at the base of the jet, leading to a hydrodynamic
jet. \citet{2015MNRAS.450.1077B} reached a similar conclusion based on
the short breakout time of the jet from the star that their model
predicted. Indeed, we find that efficient dissipation of magnetic
energy takes place at the recollimation nozzle, however it affects not
all, but about half of the magnetic energy.

\section{Discussion and Conclusions}
\label{sec:disc-concl}
In this work we present the results of a global numerical 3D MHD
stability study of relativistic Poynting flux dominated jets that
propagate in an ambient medium.  The stability of a jet largely
depends on the way it is launched and the ambient medium it interacts
with. In order for our simulations to have predictive power when it
comes to jet stability, we set them up to launch the jets in the same
way as nature does it: via the magnetised rotation of a central
compact object.
The magnetised outflow thus produced has a wide opening angle
initially. As it interacts with the ambient medium, magnetic pinch
forces collimate it into twin narrow oppositely directed
near-cylindrical jets. In this way, the structure of the jets is
established self-consistently, without any a priori assumptions about,
e.g., the jet Lorentz factor, opening angle, or magnetic pitch
angle. All of these quantities control jet stability (see
Sec.~\ref{sec:Primer}) and are determined self-consistently by the jet
launching physics. In fact, our models are characterised by just two
basic dimensionless parameters describing the astrophysical system:
the energy density of the poloidal magnetic field of the central
compact object (in units of the ambient gas density) and the slope of
the ambient gas density distribution (see
Table~\ref{tab:models_properties}).

We make a distinction between the jets that propagate in an empty,
pre-drilled funnel and the jets that have to drill their funnel for
themselves. The former, \emph{headless jets}, propagate unobstructed,
accelerate to high Lorentz factors, and do not easily show signs of
instability. The latter, \emph{headed jets}, have much larger pitch
angles (reflecting a stronger toroidal magnetic field component) and
propagate slower due to the need to expend energy on the drilling through
the ambient gas; both of these effects make headed jets unstable to
current-driven, non-axisymmetric instabilities.

As headed jets run into the ambient medium, they collimate off of
it. In fact, they recollimate and rebound, which leads to a
nozzle-like shape of the jet (see Fig.~\ref{fig:3D_jet_fieldline}). We
show that recollimation--induced--compression causes the jets to become
unstable to the internal kink mode.
This instability appears without any
perturbations imparted to the jets or the ambient gas and leads to
efficient dissipation of the toroidal magnetic field component. The
instability operates until the equipartition between thermal and
magnetic energies is reached: the thermal pressure stabilises
the jets against the further growth of the internal kink
\citep{1979SoPh...64..303H}, and the dissipation stops at plasma
$\beta=p_{\rm th}/p_{\rm mag}\sim 1$.

We find that the
dissipation rate is comparable to the growth rate of the internal kink
modes, with most of the dissipation taking place close to the
recollimation point. Our simulations do not include explicit
resistivity. Encouragingly, we find that the dissipation rate is
rather insensitive to the numerical resolution. The rate is consistent
with that found in resistive MHD simulations of internal kink instability
in the context of solar flares
\citep[e.g.][]{2009A&A...506..913H,
  2011ApJ...729..101G,2015A&A...576A..37P} and ideal relativistic MHD simulations of
steady state jets in periodic boxes
\citep[e.g.][]{2012ApJ...757...16M, 2012MNRAS.422.1436O}. It is
therefore possible that in some cases the dissipation rate can be
determined by the rate at which the macroscopic instability creates
and brings together field lines of opposite polarity and is rather
insensitive to reconnection microphysics.

We stress that whereas in our simulations jet recollimation leads to
dissipation, the nature of dissipation is different from the usual
picture of dissipation at a recollimation shock in a hydrodynamic jet
\citep{1995ApJ...449L..19G,2008Natur.452..966M,2009ApJ...699.1274B,2009MNRAS.392.1205N,2012bhae.book.....M,2012MNRAS.420L..48N,2012A&A...543A.115N,2014ApJ...787..151C}.
Even highly magnetised jets can undergo a recollimation shock provided
the jet is super-fast magnetosonic. However, not much dissipation
occurs at the shock itself, and instead the dissipation happens due to
the activation of the internal kink instability in the
post-shock/post-recollimation region: at the shock, the flow velocity
decreases and $b_\phi$ increases, both of which encourage the
development of an internal kink mode.  In the context of AGN, such a
jet recollimation can be responsible for bright features seen in the
jets at around the Bondi radius: for instance, the HST-1 component in the
M87 jet
\citep{1999ApJ...520..621B,2013ApJ...774L..21M,2015arXiv150401808H}
could be caused by the dissipation via the internal kink mode near the
recollimation point where the jet starts interacting with the
interstellar medium and the bright components emerging from and/or
passing through HST-1
might be the dissipation sites caused by the kink instability.

The jets emerge from the dissipation region with
the toroidal magnetic field component greatly weakened and the thermal
jet content much boosted by the dissipation.  This has a stabilising
effect on the internal kink mode and causes the dissipation to be
localised near the jet recollimation point. This dissipation
continues long after the jets break out of the star (see
Sec.~\ref{ssec:equipart}). Since the magnetic energy is dissipated deep inside the star, the
emerging plasma contains enough photons to account for the GRB
emission \citep{2013ApJ...764..143V}.
However, the non-thermal emission observed in many GRBs likely
requires additional dissipation to occur close to or beyond the
photosphere \citep[see, e.g.,][]{2008A&A...480..305G}.  This can
happen if the plasma cools by adiabatic expansion or
sub-photospheric emission and regains high enough magnetisation for
the internal kink mode to be revived.  It is also conceivable that if the
plasma undergoes a rapid expansion phase when exiting the star, parts
of it may attain a low enough magnetisation \citep{tch10b}, and the
internal shocks may become effective; however, it is unclear if
these parts carry sufficient amounts of energy to
power the prompt emission. Additional magnetic dissipation can be
expected if the magnetic flux threading the jets switches polarity
\citep[e.g.][]{2002A&A...391.1141D, 2005A&A...430....1G, nkt11, mu12,
  2015ApJ...805..163D}, either due to the oblique magnetic field of a
central magnetar producing a striped wind \citep{2011MNRAS.413.2031M}
or due to the accretion of matter with alternating magnetic field
polarity onto the central BH
\citep[e.g.,][]{tg14,2015MNRAS.446L..61P}. 
We find that the jet residual magnetic field content beyond the
recollimation point is sufficient to trigger the external kink
instability, which leads to large-scale helical motions and bends, 
causing the jet head to wobble (see Fig.~\ref{fig:3dextkink}). As a result, the effective
cross-section of the jets increases and the jet propagation velocity
decreases. Using an analytic model, we show that the stability of our
relativistically-hot jets to the external kink mode is controlled by
the following dimensionless parameter,
\begin{equation}
\Lambda\propto \left(\frac{L_{\rm j}}{\rho_a z_{\rm h}^2 \gamma_{\rm j}^2
    c^3}\right)^{1/6},
\notag
\end{equation}
where $\rho_a$ is the ambient density, $z_{\rm h}$ is the distance to the
jet head, and $\gamma_{\rm j}$ is the Lorentz factor of jet material (which
is larger than that of the jet head; see Sec.~\ref{sec:analytical}).  When $\Lambda$ substantially
exceeds unity, the external kink does not have sufficient time to
evolve in the jets, and the wobbling motions remain small. In this
case, the jets propagate at a similar velocity to hydrodynamic jets.
However, when $\Lambda\lesssim1$, the jet heads become unstable to
external kink modes. The jet head velocity significantly decreases and
becomes much smaller than that of hydrodynamic jets. Using our
analytic model, we show the stability of the jets and the propagation
velocity of their heads are closely related to the profile of the
external density. In ambient density profiles steeper than
$\rho \propto r^{-2}$, the jets accelerate and become more stable with
increasing distance. In flatter profiles, however, the jets decelerate
and become exceedingly unstable.  We expect that eventually the head
velocity becomes comparable to the expansion velocity of the shocked
external gas, which then outruns the jets. This may cause the jets to
stall.

The concept of a stalled jet can explain the difference in the
morphology of FRI and FRII radio galaxy jets
\citep{1974MNRAS.167P..31F}. 
The jets of FRI galaxies usually extend
to several tens of kpc, are engulfed in a massive cocoon, and shine
along the entire body. FRII galaxy jets, on the other hand, are much
longer on average ($\simgreat 100$~kpc), maintain straight course and
have a well-defined, bright, hot spot at the jet head. These
differences can be explained via the effects of the external kink
instability. In the flat density profiles of galaxy cluster cores, the
jets decelerate as they propagate outwards. Therefore, a minimal power
exists above which they are able to break out of these cores. We
estimate this minimal power to be $L^{\rm FR}_{\rm min}\sim4\times10^{45}$ erg~s$^{-1}$, adopting typical properties of radio galaxies (see
Sec.~\ref{sec:astrophysical} and
Tab.~\ref{tab:models_properties}). Jets with a lower power become
external--kink--unstable inside the core. They develop wobbling
motions which increase their effective cross-section and cause them to
slow down until they are overrun by their cocoons that continuously expand. The increased wobbling motions can also trigger dissipation
along the jets that leads to radiation, making the jets shine and appear
as FRI jets.
A higher luminosity jet will break out of the flat density core before
becoming highly kink-unstable and will continue to accelerate in the
steep ambient density profile that surrounds the galaxy core. Such a
jet will remain stable and will appear as an FRII jet.

In the GRB case, the density profile of the stellar envelope is so
steep that the jets become exceedingly more stable as they propagate
outward. We find that typical GRB jets are marginally stable and
propagate at a velocity $\simless0.5 c$, similar to the velocity of
hydrodynamic jets (see Fig.~\ref{fig:sim_vs_model}). They break out of the star in about
$\sim10$~seconds. \citet{2015MNRAS.450.1077B} concluded that Poynting
flux dominated jets need to dissipate their energy close to the source
in order for the breakout time to be consistent with observations. We
show here that such a dissipation mechanism indeed exists near the
recollimation point of the jet. The breakout time agrees with that
inferred from the distribution of GRB duration times
\citep{2012ApJ...749..110B,2015MNRAS.450.1077B}. The similarity
between the propagation velocities of hydrodynamic and our
relativistically-hot MHD jets suggests that it is difficult if not
impossible to distinguish between magnetic and hydrodynamic energy
injection mechanisms in the jets based on the observed breakout time
alone.

We find that if the power of a (single) GRB jet drops below
$\sim10^{47}$~erg~s$^{-1}$, which corresponds to an isotropic
equivalent luminosity of $\sim2.6\times10^{49} (\theta/7^\circ)^{-2}$,
the jet becomes so strongly unstable to the external kink mode inside
the star that its propagation velocity significantly drops (see Sec.~\ref{sec:astrophysical}). This
can lead to a situation where the GRB engine shuts off before the jet
is able to breakout of the star. In this case, only the cocoon breaks
out of the star, and feasibly leads to a different type of burst,
which may resemble a low-luminosity GRB \citep{1998Natur.395..663K,
  2001ApJ...550..410M, 2001ApJ...551..946T, 2007ApJ...664.1026W,
  2007ApJ...667..351W, 2010ApJ...716..781K, 2011ApJ...739L..55B,
  2012ApJ...747...88N,2015ApJ...807..172N}.

We note that whereas our simulations are carried out for a neutron star as the
central compact object, our results are applicable to the case of an
accreting black hole. This is because both black holes and neutron
stars launch the jets via the azimuthal winding of poloidal magnetic
field lines, as we discussed previously. The difference is that the
field line winding in the case of the neutron star is due to the
rotation of an actual, physical surface with magnetic field lines
frozen into it, and in the black hole case it is due to the rotation
of the space-time~\citep{2015ASSL..414...45T}. Note, however, that the
presence of a black hole accretion disc, which we do not include in
this work, may change the conditions at the base of the jet and affect
the jet structure. For instance, a disc wind may alter the confining
pressure profile of the jet.
These effects will be studied elsewhere.

In this work, we only studied jet propagation for 3 orders of
magnitude in distance. This is short by about an order of magnitude in
distance of the actual stellar surface.  Carrying out longer-term
simulations that extend up to and beyond the stellar surface are
important for verifying the stability of the jet out to larger
distances and testing our analytic models in this new regime. We have
not studied the processes that occur in the jets upon their breakout of the
star. The breakout can be accompanied with substantial
acceleration \citep{tch10b} and the loss of causal contact across the
jet \citep{kvk09}, both of which can affect the development of the
instability. The simulations of jet breakout are important also in
order to compute a first-principles structure of the jet outside of
the star to improve the GRB prompt and afterglow emission modelling.

The main finding of this work is the efficient dissipation of the jet
magnetic field near the recollimation point via the internal kink
instability. Though the dissipation rate is presently difficult to
measure quantitatively due to the inherent time variability and potential
sensitivity to the microphysics,
the outcome of the internal kink instability---a hot jet with thermal
and magnetic energies in equipartition---is robust. It was obtained by
multiple groups in various contexts using different numerical methods.
Thus, the jet material that emerges from the star is both
relativistically-hot and relativistically-magnetised.  
We conclude
that the internal kink instability leads to jets that have enough
magnetic and thermal energy to accelerate to the highly relativistic
Lorentz factors inferred for GRBs, and carry a substantial thermal
energy flux, which can be used to power the prompt emission.

\section*{acknowledgments} 
We thank R.~Barniol Duran, E.~Blackman, D.~Giannios, H.~Li,
Y.~Lyu\-bar\-s\-ky, J.~McKinney, R.~Narayan,
K.~Parfrey, A.~Philippov, S.~Phinney, E.~Quataert, F.~Ryde, and A.~Spitkovsky
for discussions that helped us to improve this work. O.B.\ acknowledges
support from the Lyman Spitzer Jr.\ Fellowship, awarded by the
Department of Astrophysical Sciences at Princeton University and the
Max-Planck/Princeton Center for Plasma Physics which facilitated this
work.  A.T.\ was supported by NASA through Einstein Postdoctoral
Fellowship grant number PF3-140115 awarded by the Chandra Xray Center,
which is operated by the Smithsonian Astrophysical Observatory for
NASA under contract NAS8-03060, and NSF through an XSEDE computational
time allocation TG-AST100040 on NICS Kraken, Nautilus, TACC Stampede,
Maverick, and Ranch. The simulations presented in this work also used
computational resources supported by the PICSciE-OIT High Performance
Computing Center and Visualization Laboratory, and the Savio cluster
provided by UCB.

\appendix

\section{Analytic model of the jet}
\label{sec:analytic-model-jet}
In this Appendix we calculate the jet radius following the calculation
method of \citet{2011ApJ...740..100B}, with some minor modifications
to fit the specific model of the magnetised jet. As illustrated in
Fig.~\ref{fig:jet_cocoon_model}, we assume a
cylinder--shaped jet surrounded by a cylindrical cocoon propagating in
a cold medium. The jet propagates in the $z-$direction and injects
energy into the cocoon at the outermost parts jet of the jet, or at the \emph{jet
  head}. As a result the cocoon becomes pressurised and expands
sideways into the ambient medium, as seen in Figure
\ref{fig:jet_cocoon_model}.  The
calculation here is relevant for an external medium density with a
power-law profile: $\rho_a\propto z^{-\alpha}$. The power-law
assumption implies that all the quantities in the problem have a
power-law dependency on $z$ as well, which makes it easier to
integrate them.

According to eq. \eqref{eq:pc0} the cocoon pressure is
 \begin{equation}\label{pc1_app}
  p_c=\frac{E_c}{3V_c}=\frac{\int L_{\rm j}(\beta_{\rm j}-\beta_{\rm h})dt}{3\pi\int{\beta_{\rm h}dt}\left(\int{\beta_{\rm c}dt}\right)^2},
\end{equation}
where $\beta_{\rm h}$ is the velocity of the jet head and
$\beta_{\rm j}$ is the velocity of the jet material that lies below the head
and that carries the jet energy to the head. $L_{\rm j}$ is the power, or
luminosity, injected into the jet by the engine located at the base of
the jet and the term $L_{\rm j}(\beta_{\rm j}-\beta_{\rm h})$ represents the
fraction of that luminosity that reaches the jet head and is injected into the
cocoon. $\beta_{\rm c}$ is the average expansion velocity of the cocoon in
the transverse direction. Since the cocoon expands into a cold gas, it
drives a shock into the ambient medium. We can approximate the
expansion velocity of the cocoon as the velocity of that shock:
$\beta_{\rm c}=\sqrt{p_c/\bar{\rho}_ac^2}$, where $\bar{\rho}_a$ is the
average ambient density along the shock's upstream. Substituting this expression
in eq.~\eqref{pc1_app}, we get (up to integration constants, which can be
calculated posteriorly):
\begin{equation}\label{pc2_app}
  p_c\approx\sqrt{\frac{\bar{\rho}_a c^2L_{\rm j}(\beta_{\rm j}-\beta_{\rm h})}{3\pi{\beta_{\rm h}}t^2}},
\end{equation}
Following \citet{2003MNRAS.345..575M,2011ApJ...740..100B} we define a dimensionless parameter
\begin{equation}\label{Ltilde_app}
  \tilde{L}\equiv\frac{L_{\rm j}}{\pi R_{\rm j}^2\rho_a c^3}=\fracb{\beta_{\rm j}-\beta_{\rm h}}{\beta_{\rm h}}^2,
\end{equation}
where $R_{\rm j}$ is the cylindrical radius of the jet.
Substituting eq.~\eqref{Ltilde_app} in eq.~\eqref{pc2_app}, and defining $t\simeq{z_{\rm
    h}}/{\beta_{\rm h} c}$, where $z_{\rm h}$ is the distance to the jet head, 
we get (up to a constant)
\begin{equation}\label{pc3_app}
  p_c\simeq\left(\frac{\bar{\rho}L_{\rm j}}{\pi c}\right)^{1/2}\tilde{L}^{-1/4}t^{-1}=\left(\frac{\bar{\rho}_aL_{\rm j}c}{\pi z_{\rm h}^2}\right)^{1/2}\fracb{L_{\rm j}}{\pi R_{\rm j}^2\rho_a c^3}^{1/4}
\end{equation} 

The cocoon maintains a pressure balance with the jet along the jet edge (at $R_{\rm j}$). We can calculate $p_j(R_{\rm j})$ the equation for the total jet power:
 \begin{equation}
L_{\rm j}=2\pi\int_0^{R_{\rm j}}
\frac{\Gamma}{\Gamma-1}p_j\gamma_{\rm j}^2\beta_{\rm j}cRdR,
\label{Lj_app}
\end{equation}
 where the $p_j$ is given in eq. \eqref{eq:Pj} and 
 $\Gamma$ is the adiabatic index. To calculate $\Gamma$ we can write the total jet pressure as the sum of the thermal pressure and the magnetic pressure:
 \begin{equation}
   p_j=(\Gamma_{\rm th}-1)u_{\rm th} + (\Gamma_{\rm mag}-1)u_{\rm mag}=\frac{1}{3}u_{\rm th}+u_{\rm mag},
   \label{pj_Gamma_app}
 \end{equation}
 where $\Gamma_{\rm th}$ ($\Gamma_{\rm mag}$) and $u_{\rm th}$ ($u_{\rm mag}$) are the adiabatic index and the energy density of the gas (magnetic field) respectively. In Sec. \ref{ssec:equipart} we show that above the dissipation region the thermal and magnetic pressures are in equipartition, implying that $u_{\rm th}=3u_{\rm mag}$. Substituting that in eq. \eqref{pj_Gamma_app} we get that 
 \begin{equation}
   p_j=\frac{1}{2}(u_{\rm th}+u_{\rm mag}),
 \end{equation}
 This gives an adiabatic index inside the jet of
 \begin{equation}
   \Gamma=\frac{p_j}{u_{\rm th}+u_{\rm mag}}+1=\frac{3}{2}.
   \label{Gamma_app}
 \end{equation}
Integrating eq. \eqref{Lj_app} with $\Gamma$ from \eqref{Gamma_app} we get that the jet pressure at $R_{\rm j}$ is
\begin{equation}\label{pj_app}
  p_j(R_{\rm j})\simeq\frac{2L_{\rm j}}{9\pi\gamma_{\rm j}^2\beta_{\rm j}R_{\rm j}^2c}.
\end{equation} 

The balance between the jet pressure and the cocoon pressure at the jet edge implies, $p_c(R_{\rm j})=p_j(R_J)$.  Using eqn.~\eqref{pc3_app} and \eqref{pj_app} we get:
\begin{equation}
  R_{\rm j}\gamma_{\rm j}=\fracb{L_{\rm j}z_{\rm h}^4}{\rho_ac^3\gamma_{\rm j}^2\beta_{\rm j}^3}^{1/6}\sqrt{\frac{2}{9\pi\beta_{\rm j}}}\left(\frac{3\pi}{2}\frac{(5-\alpha)(3-\alpha)}{9}\right)^{1/3},
\end{equation}
where we included the numerical factor obtained from the integrations
and defined $\rho_a$ as the ambient density at $z_{\rm h}$.
To calculate the numerical factor we used the following relations \citep[see][appendix B]{2011ApJ...740..100B}:
$\bar{\rho}_a=\int\rho_adV/V=\frac{3}{3-\alpha}\rho_a$;  $z_{\rm h}=\int\beta_{\rm h}t=\frac{5-\alpha}{3}\beta_{\rm h}dt$ and $R_{\rm c}=\int\beta_{\rm c}dt=\frac{5-\alpha}{3}\beta_{\rm c}dt$.

\label{lastpage}
\end{document}